\documentclass[reprint,aps,prd,showpacs,eqsecnum,twocolumn,superscriptaddress,nofootinbib]{revtex4-1}
\usepackage{amsmath,amssymb,graphicx,color,ulem,multirow, inputenc}
\usepackage{hyperref,ulem,url}
\hypersetup{colorlinks=true}

\begin{document}

\title{Towards high-precision inspiral gravitational waveforms from binary neutron star mergers in numerical relativity} 

\author{Kenta Kiuchi}
\affiliation{Max Planck Institute for Gravitational Physics (Albert Einstein Institute), Am M\"{u}hlenberg, Potsdam-Golm, 14476, Germany}
\affiliation{Center for Gravitational Physics and Quantum Information, Yukawa Institute for Theoretical Physics, Kyoto University, Kyoto 606-8502, Japan}

\date{\today}

\begin{abstract}
We report the performance of a newly implemented fourth-order accurate finite-volume HLLC Riemann solver in the adaptive-mesh-refinement numerical relativity code {\tt SACRA-MPI}. First, we validate our implementation in one-dimensional special relativistic hydrodynamics tests, i.e., a simple wave and shock tube test, which have analytic solutions. We demonstrate that the fourth-order convergence is achieved for the smooth flow, which cannot be achieved in our original second-order accurate finite-volume Riemann solver. We also show that our new solver is robust for the strong shock wave emergence problem. Second, we validate the implementation in a dynamical spacetime by demonstrating that {\tt SACRA-MPI} perfectly preserves the $\pi$-symmetry without imposing the $\pi$-symmetry in a short-term ($\sim 20~{\rm ms}$ in the inspiral and subsequent post-merger phase) non-spinning equal-mass binary neutron star merger simulations. Finally, we quantify the accuracy of $\approx 28$ cycles inspiral gravitational waveforms from binary neutron star mergers by conducting a resolution study with $\approx 78, 94$, $118$, and $135$ m. We find that the fourth-order accurate Riemann solver achieves the convergence order $\approx 2.1\pm{0.05}$--$2.4\pm{0.27}$, i.e., slightly evolving with time, in the inspiral gravitational wave phase, while the second-order accurate Riemann solver achieves the convergence order $\approx 2.0\pm{0.5}$. The residual phase error towards the continuum limit at the merger is $0.27\pm 0.07$ rad and $0.58\pm 0.22$ rad out of a total phase of $\approx 176$ rad, respectively, for the fourth- and second-order accurate Riemann solver. 

\end{abstract}

\maketitle

\section{Introduction}
The first detection of gravitational waves from the binary neutron star (BNS) merger GW170817 proved that the unknown neutron star equation of state (EOS) is constrained by measuring the tidal deformability during the inspiral phase~\cite{LIGOScientific:2017vwq}. The LIGO-VIRGO-KAGRA collaborations and subsequent independent works constrained the tidal deformability to $\tilde{\Lambda}\lesssim 700$~\cite{LIGOScientific:2017vwq,De:2018uhw,LIGOScientific:2018hze,Most:2018hfd,Capano:2019eae,Chatziioannou:2020pqz}, which already excludes some extreme NS EOSs. However, it is still on the way to pin down the NS EOS~\cite{Chatziioannou:2020pqz}. The next generation gravitational-wave detectors, the Einstein Telescope (ET)~\cite{Punturo:2010zz} and Cosmic Explorer (CE)~\cite{Reitze:2019iox}, will give more stringent constraints on the NS EOS by measuring $O(10^7)$ BNS merger events per year. 

Theoretical gravitational waveform templates are an essential tool to measure the tidal deformability imprinted in gravitational waves from BNS mergers~\cite{Flanagan:2007ix,Hinderer:2007mb,Damour:2009vw}. The state-of-the-art waveform modeling is categorized into two families: Effective One Body family~\cite{Buonanno:1998gg,Damour:2001tu,Nagar:2018zoe,Haberland:2025luz,Albanesi:2025txj} and Phenom family~\cite{Ajith:2009bn,Hannam:2013oca,Khan:2018fmp,Khan:2015jqa,Pratten:2020fqn}, which were originally invented for the waveform modeling for binary black hole (BBH) mergers. On top of the BBH baseline, the NS matter sector is based on analytic models up to 7.5 Post-Newtonian order~\cite{Vines:2011ud,Damour:2012yf,Henry:2020ski,Mandal:2024iug} or semi-analytic models informed by high-precision numerical relativity (NR) simulations of BNS mergers~\cite{Abac:2023ujg,Dietrich:2019kaq,Dietrich:2017aum,Dietrich:2018uni,Kawaguchi:2018gvj,Narikawa:2019xng}. However, the systematic errors among the state-of-the-art BNS waveform models are not small enough to sufficiently suppress the bias in the parameter estimation of gravitational waves in the ET and CE era~\cite{Punturo:2010zz,Reitze:2019iox}. 

NR is a chosen way to build a theoretical waveform template in the late inspiral phase of BNS mergers because (i) the NSs undergo significant tidal deformation and (ii) any analytic tools break down in this phase. Since 2010, the NR community started to make an effort to derive gravitational waveforms from BNS mergers whose quality should be fine enough to validate/calibrate the semi-analytical waveform models~\cite{Kiuchi:2017pte,Kiuchi:2019kzt,Kawaguchi:2018gvj,Hotokezaka:2013mm,Hotokezaka:2015xka,Hotokezaka:2016bzh,Narikawa:2018yzt,Narikawa:2019xng,Kuan:2024jnw,Kuan:2025bzu,Dietrich:2017aum,Dietrich:2018phi,Dietrich:2018uni,Dietrich:2019kaq,Dietrich:2016hky,Dietrich:2016lyp,Dietrich:2017feu,Bernuzzi:2011aq,Bernuzzi:2012ci,Bernuzzi:2013rza,Bernuzzi:2014kca,Bernuzzi:2014owa,Foucart:2018lhe,Haas:2016cop,Radice:2013hxh,Radice:2013xpa,Baiotti:2010xh,Gamba:2020wgg}. 

Among a couple of the error budgets in numerically derived gravitational waveforms, the truncation error due to the finite resolution is the most severe one. It implies that high-precision BNS merger simulations are necessary since numerical dissipation stemming from an approximate treatment of the partial differential equation, e.g., the finite difference, could deteriorate numerical solutions, resulting in a spurious acceleration of binary orbital decay. In other words, the gravitational wave phase error due to the numerical dissipation should be sufficiently suppressed compared to the gravitational wave phase shift due to the tidal deformation. Therefore, a higher-order scheme, particularly for a {\it hydrodynamics solver}, is an essential ingredient to achieve this requirement. To date, a couple of existing NR codes, {\tt WhiskyTHC}~\cite{Radice:2013hxh}, {\tt BAM}~\cite{Bernuzzi:2016pie}, {\tt FIL}~\cite{Most:2019kfe}, and {\tt SpEC}~\cite{Foucart:2020xkt} implement a higher-order {\it finite-difference} scheme beyond the second order for relativistic (magneto)hydrodynamics to simulate BNS mergers (see {\tt SpECTRE}~\cite{Deppe:2024ckt} and {\tt Nmesh}~\cite{Tichy:2022hpa} for a discontinuous Galerkin method) (see also the waveform comparison among these NR codes in Ref.~\cite{Hamilton:2024ziw}). 

However, there is no NR code implementing a higher-order {\it finite-volume} method beyond the second order for relativistic hydrodynamics. Since the finite-volume method could be preferred over 
the finite difference scheme to simulate astrophysical fluid dynamics in the sense that it fully relies on solving the so-called Riemann problem to capture the discontinuous solution, such as the shock wave, it is desired to implement a higher-order finite-volume Riemann solver in NR codes. Also, the quality of the Riemann solver should be improved. References~\cite{Kiuchi:2022,Kuan:2024jnw} demonstrate that the Harten-Lax-van Leer-contact (HLLC) Riemann solver~\cite{Mignone:2005ft} is superior to the Harten-Lax-van Leer-Einfeldt (HLLE) solver, which most existing NR codes implement. Particularly, not only the quantitative improvement of the non-spinning BNS merger dynamics~\cite{Kiuchi:2022} but also the qualitative improvement to simulate the $f$-mode resonance in the rapidly anti-spinning BNS merger are reported~\cite{Kuan:2024jnw}.

In this paper, we report an implementation of the fourth-order accurate finite-volume HLLC Riemann solver in the NR code {\tt SACRA-MPI}~\cite{Yamamoto:2008js,Kiuchi:2017zzg} following the novel strategy~\cite{Berta:2023zmr}. The paper is organized as follows. In Sec.~\ref{sec:implementation}, we describe an implementation and its validation of the fourth-order accurate finite-volume Riemann solver in our NR code {\tt SACRA-MPI}. Section~\ref{sec:result} is devoted to quantifying the performance of the new fourth-order Riemann solver in BNS merger simulations. In particular, we conduct quantitative comparisons to the second-order finite-volume Riemann solver. 
Discussion is given in Sec.~\ref{sec:discussion}. 
In Sec.~\ref {sec:conclusion}, we summarize this paper. Throughout the paper, we use the geometrical unit $c=G=1$, unless otherwise stated, where $c$ and $G$ denote the speed of light and the gravitational constant, respectively. 

\section{Fourth-order accurate finite-volume Riemann solver}\label{sec:implementation}
\subsection{Implementation}
We implement a fourth-order HLLC Riemann solver in the framework of the finite-volume method. The readers may refer to Ref.~\cite{Kiuchi:2022} for the derivation of the finite-volume method and a tetrad transformation to a locally Minkowski spacetime in a curved spacetime. In general, the higher-order Riemann solver beyond the second order is not straightforward in the finite-volume method because of the primitive-conservative conversion. We note that the higher-order {\it finite difference} Riemann solver is straightforward~\cite{Radice:2013hxh,Bernuzzi:2016pie,Most:2019kfe,Foucart:2020xkt}.

Assuming a smooth distribution inside a cell, a volume-averaged variable, $\bar{Q}$, is expressed by the point-wise (PW) variable, $Q^\text{PW}$, with the fourth-order accuracy: 
\begin{widetext}
\begin{align}
    \bar{Q} _j \equiv \frac{1}{\Delta x} \int^{x_j+\Delta x/2}_{x_j-\Delta x/2} Q(x) {\mathrm d}x = Q^\text{PW}_j + \frac{1}{24}\partial_{xx} Q^\text{PW}|_{x=x_j}\Delta x^2 + O(\Delta x^4), \label{eq:PWtoFV}
\end{align}
\end{widetext}
where we consider the cell $x \in [x_j-\frac{\Delta x}{2},x_j+\frac{\Delta x}{2}]$ with the cell width $\Delta x$. However, this equation implies a volume-averaged conserved variable, e.g, a momentum $\overline{\rho v}$, can not be expressed by the multiplication of volume-averaged primitive variables, e.g., the rest-mass density $\bar{\rho}$ and the velocity $\bar{v}$ at the fourth-order accuracy. Namely, $\overline{\rho v} \ne \bar{\rho} \times \bar{v}+O(\Delta x^4)$ because of the higher-order term proportional to $\Delta x^2$ in the right-hand side of Eq.~(\ref{eq:PWtoFV}). Therefore, the primitive-conservative conversion for the {\it volume-averaged} variable automatically degrades the accuracy down to the second order. Reference~\cite{Berta:2023zmr} proposes a novel method to maintain the fourth-order accuracy in the finite-volume method. 

Their methodology is summarized as follows. As a first step, we construct a fourth-order accurate point-wise conserved variable from the volume-averaged conserved variable by
\begin{widetext}
\begin{align}
Q^\text{PW}_j &= \bar{Q}_j - \frac{\theta_c}{24}\partial_{xx} \bar{Q}_j\Delta x^2 + O(\Delta x^4) =\bar{Q}_j - \frac{\theta_c}{24}\left( \bar{Q}_{j+1} - 2 \bar{Q}_j + \bar{Q}_{j-1} \right) + O(\Delta x^4),\label{eq:FVtoPW}
\end{align} 
\end{widetext}
where we introduce a shock detector, $\theta_c$, which is reduced to zero at the discontinuity to suppress undesired oscillations. For a smooth profile, $\theta_c=1$. 

As a second step, we perform a primitive recovery for the fourth-order accurate point-wise conserved variable, $Q^\text{PW}$, to obtain the fourth-order accurate point-wise primitive variable, $P^\text{PW}$. Because the primitive recovery is done for the point-wise variable, it maintains the fourth-order accuracy. We should note that Eq.~(\ref{eq:FVtoPW}) cannot be applied to the point-wise primitive variable to maintain consistency between the point-wise conserved and primitive variables.

As a third step, we apply a higher-order reconstruction scheme to the fourth-order accurate point-wise primitive variable, $P^\text{PW}$, to obtain a left and right state at the cell interface, $P^\text{PW}_{L/R}$, and feed them to a Riemann solver to calculate a numerical flux. 
Again, the cell reconstruction for the point-wise variable maintains the fourth-order accuracy as long as the cell reconstruction accuracy is higher than fourth order.

Practically, we implement the MP5 scheme~\cite{Suresh:1997} for the cell reconstruction, and the HLLC solver for the Riemann solver~\cite{Mignone:2005ft,Kiuchi:2022}. We also estimate the shock detector $\theta_c$ by using the five stencils as proposed in Ref.~\cite{Berta:2023zmr} 
to distinguish the discontinuity and extrema. 
Specifically, $\theta_c$ is estimated by
\begin{align}
\theta_c = \left\{\begin{array}{cc}
1 & \text{if }\eta_c < \eta_d, \\
0 & \text{otherwise,}
\end{array}
\right.
\end{align}
where $\eta_c=\sqrt{\eta_{c,x}^2+\eta_{c,y}^2+\eta_{c,z}^2}$ and $\eta_d$ is a threshold parameter which is set to be $0.005$ in our implementation. For example, $\eta_{c,x}$ at $x_j$ is estimated by $\max(\eta^o_{c,x},\eta^e_{c,x})$ where
\begin{widetext}
\begin{align}
\eta^o_{c,x} = \frac{|\delta^{(3)} \bar{Q}_c|}{|\bar{Q}_\text{c,ref}| + |\delta^{(1)} \bar{Q}_c|+|\delta^{(3)} \bar{Q}_c|+\epsilon},~
\eta^e_{c,x} = \frac{|\delta^{(4)} \bar{Q}_c|}{|\bar{Q}_\text{c,ref}| + |\delta^{(2)} \bar{Q}_c|+|\delta^{(4)} \bar{Q}_c|+\epsilon},
\end{align}
where
\begin{align}
&\delta^{(1)}\bar{Q}_c = \frac{1}{2} \left(\bar{Q}_{j+1}-\bar{Q}_{j-1}\right),~\delta^{(2)}\bar{Q}_c = \left(\bar{Q}_{j+1}-2\bar{Q}_j+\bar{Q}_{j-1}\right),~\delta^{(3)}\bar{Q}_c = \frac{1}{2}\left(\bar{Q}_{j+2}-2\bar{Q}_{j+1}+2\bar{Q}_{j-1}-\bar{Q}_{j-2}\right),\nonumber\\
&\delta^{(4)}\bar{Q}_c = \bar{Q}_{j+2}-4\bar{Q}_{j+1}+6\bar{Q}_j-4\bar{Q}_{j-1}+\bar{Q}_{j-2},
\end{align}
\end{widetext}
and we set $\epsilon=10^{-15}$ to avoid the denominator from being zero. We compute $\eta^{o/e}_{c,x}$ with the volume-averaged rest-mass density, pressure, and three velocities, respectively. For the reference value $\bar{Q}_\text{c,ref}$, we employ the volume-averaged rest-mass density and pressure, while we employ the iso-thermal sound speed $\sqrt{\bar{P}/\bar{\rho}}$ for the three velocity components. 

If the shock is detected, i.e., $\theta_c=0$ in a cell, we switch to a lower-order reconstruction scheme (the third-order accurate PPM scheme in our case) in this trouble cell. 

For the multidimensional case, we calculate the fourth-order accurate numerical flux at the cell interface in the x-direction, $\hat{f}_{j+1/2,k,l}$, by
\begin{widetext}
\begin{align*}
 &\hat{f}_{j+1/2,k,l} = \tilde{f}_{j+1/2,k,l} + \frac{\theta_c}{24} 
 \left( \partial_{yy} \tilde{f}_{j+1/2,k,l} \Delta y^2 + \partial_{zz} \tilde{f}_{j+1/2,k,l} \Delta z^2 \right) + O(\Delta x^4,\Delta y^4,\Delta z^4)\nonumber\\
 & = \tilde{f}_{j+1/2,k,l} + \frac{\theta_c}{24} \Big(-4 \tilde{f}_{j+1/2,k,l}+\tilde{f}_{j+1/2,k+1,l} +\tilde{f}_{j+1/2,k-1,l}
+ \tilde{f}_{j+1/2,k,l+1}+\tilde{f}_{j+1/2,k,l-1}
 \Big)  + O(\Delta x^4,\Delta y^4,\Delta z^4), 
\end{align*}
\end{widetext}
where $\tilde{f}_{j+1/2,k,l}$ is the numerical flux calculated by the fourth-order accurate Riemann solver in the x-direction. The higher-order terms proportional to $\Delta y^2$ and $\Delta z^2$ in the right-hand side are a consequence of the numerical flux distribution on the cell surface with $x=x_j+\Delta x/2$, $y\in[y_k-\Delta y/2,y_k+\Delta y/2]$, and $z\in [z_l-\Delta z/2,z_l+\Delta z/2]$ (see also Eq.~(\ref{eq:PWtoFV})).  
We again introduce the shock detector in front of the higher-order term. 

As a final step, we update the volume-averaged conserved variable with the numerical fluxes $\hat{f}_{j+1/2,k,l}$ and execute the primitive recovery for the volume-averaged variables. In this way, the fourth-order accuracy is maintained in the finite-volume method. 

Finally, we remark that the fourth-order accurate Riemann solver requires an additional primitive recovery (step 2) compared to the second-order accurate Riemann solver. From a computational cost point of view, it is a drawback since the primitive recovery is a time-consuming part in relativistic hydrodynamics. However, we found that this drawback is compensated in a BNS merger at least for the analytical EOS (see Sec.~\ref{sec:result}).

Before closing this subsection, we should mention the treatment of the source term in the Einstein equation and the hydrodynamics equation. We employ the point-wise variables for the source term in the Einstein equation. In a curved spacetime, the source term also appears in the relativistic Euler and energy equations. Since it is composed of a coupling of the geometrical and hydrodynamical variables, it is non-trivial to obtain the fourth-order accurate volume-averaged source term (see Eq.~(3.15) in Ref.~\cite{Kiuchi:2022}). In this paper, we employ the volume-averaged variables for the source term in the hydrodynamics equation, assuming a constant profile for the determinant of the three metric and that for the spatial derivative of the other geometrical variables inside a cell. 
However, there is room for improvement or for constructing a true fourth-order source term.

\subsection{Validation in flat spacetime}
\subsubsection{Problem 1: Simple wave in relativistic hydrodynamics}
\begin{figure*}
    \centering
    \includegraphics[scale=0.35]{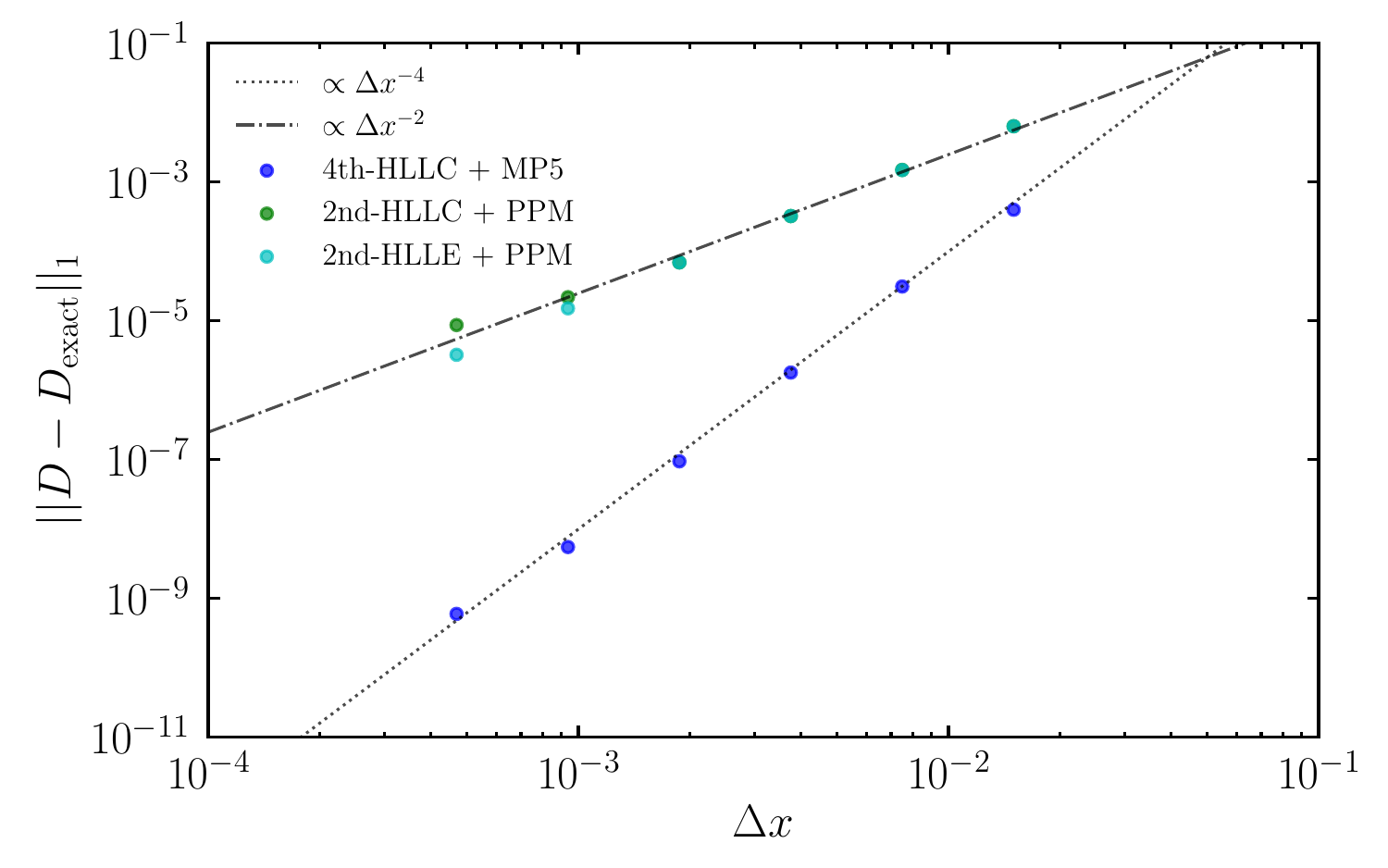}
    \includegraphics[scale=0.35]{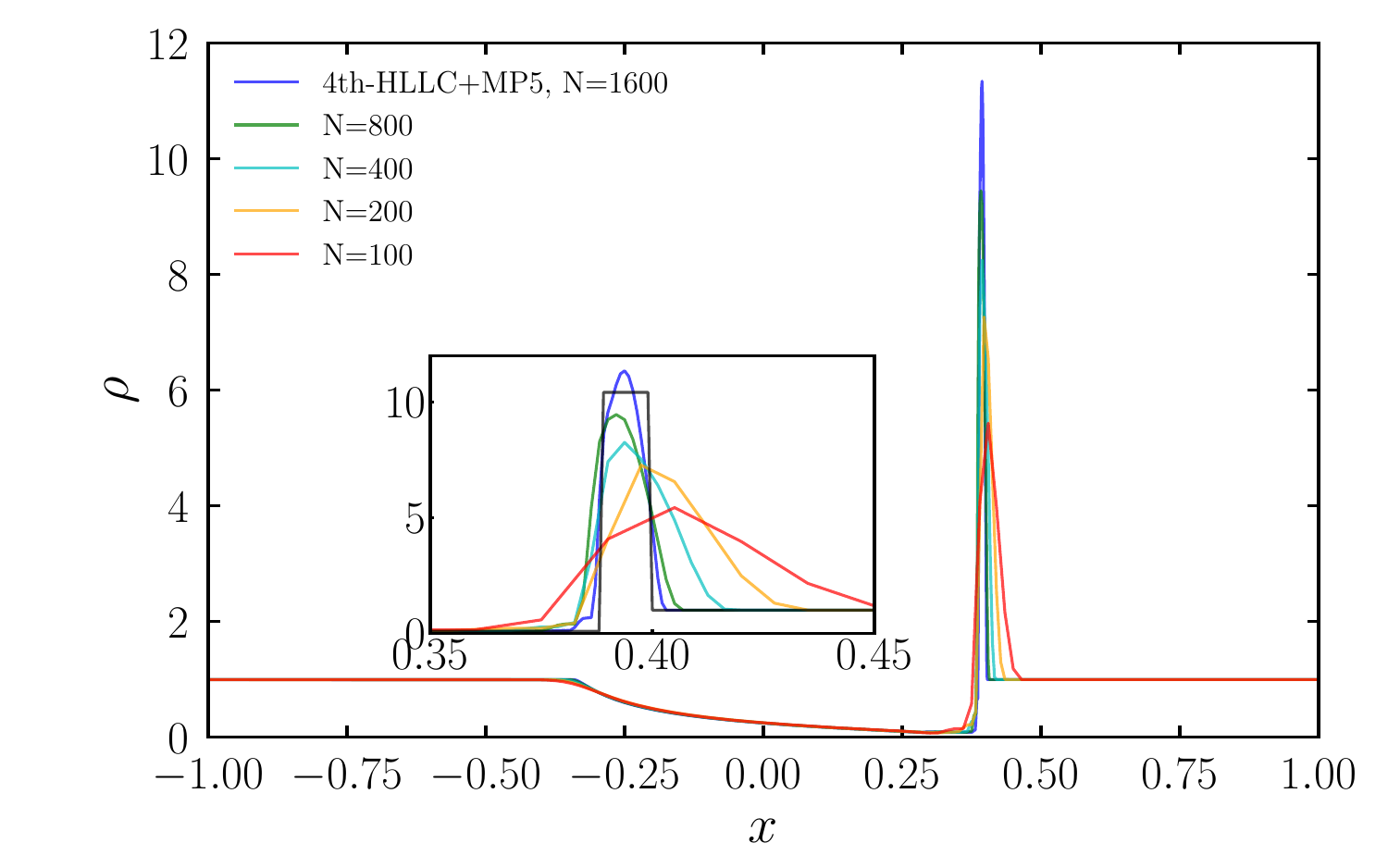}
    \caption{(Left) The L1 norm of the error for the conserved mass density in the simple wave test. We estimate the error from the exact solution at $t=0.60$. The blue dots are the result of the fourth-order accurate HLLC Riemann solver combined with the MP5 cell reconstruction scheme. The green (cyan) dots are the result of the second-order accurate HLLC (HLLE) Riemann solver with the third-order accurate PPM cell reconstruction. 
    The black dotted (dot-dashed) line presents an expected fourth-order (second-order) accuracy. (Right) The rest-mass density profile in the shock tube problem. The inset shows a close-up of the shock as well as the contact discontinuity. The black-solid curve denotes the exact solution.}
    \label{fig:Simple_Wave}
\end{figure*}

We validate the new fourth-order accurate Riemann solver in a relativistic simple wave test~\cite{Liang:1977,Bernuzzi:2016pie}. We set the uniform and static flow with $\rho=1$ and $v^x=0$ as a reference state. On top of it, we prescribe the velocity perturbation by
\begin{align}
\delta v^x = a \Theta \left(|x|-X\right) \sin^6 \left(\frac{\pi}{2}\left(\frac{x}{X}-1\right)\right),
\end{align}
where $\Theta$ is the Heaviside function. We set $a=0.5$, and $X=0.3$. We also assume the polytropic EOS with the polytropic constant $K=100$ and index $\Gamma=5/3$. With this particular setup, the shock will appear at $t\simeq 0.63$~\cite{Liang:1977,Bernuzzi:2016pie}. Therefore, we terminate a simulation at $t=0.60$ before the shock appears. The numerical domain is $x\in [-1.5,1.5]$ and the $x$-coordinate is set to $x_j = j \Delta x $ with $j \in [-N,N]$ and $\Delta x=1.5/N$. We vary $N=3200,1600,800,400,200,100$ to check the convergence. The left panel of Fig.~\ref{fig:Simple_Wave} depicts the L1 norm of the error from the exact solution.  
It shows an almost perfect fourth-order convergence~\footnote{Since the exact solution is a point-wise value, it is necessary to add the higher-order correction $\propto \Delta x^2$ in Eq.~(\ref{eq:PWtoFV}) at the initialization of the volume-averaged conserved variable. Otherwise, the fourth-order convergence is never achieved.}. 
The plot also shows  the original second-order accurate HLLC/HLLE solver with the third-order PPM reconstruction. The errors decrease with the second-order accuracy. These findings validate our implementation.

\subsubsection{Problem 2: Shock tube problem}
The second validation is the case where a shock wave appears. We employ the shock tube problem (Problem HD4 in Ref.~\cite{Kiuchi:2022}) where a strong shock wave emerges. The initial condition is $\rho_{L/R}=1$, $v^x_{L/R}=0$, and $P_{L/R}=10^3/10^{-2}$. The EOS is the $\Gamma$-law with $\Gamma=5/3$. 
The simulation domain is $x\in[-1,1]$, and the grid structure is set to $x_j=j\Delta x$ with $j \in [-N,N]$ and $\Delta x=1/N$. We vary $N=1600,800,400,200,100$. We terminate the simulation at $t=0.4$. The right-hand panel of Fig.~\ref{fig:Simple_Wave} shows the profile for the rest-mass density on top of the exact solution. On the one hand, by improving the resolution, the shock and contact discontinuity are sharply resolved (see the inset). On the other hand, unphysical oscillations do not appear, implying that the shock detector $\theta_c$ properly works. This result validates our implementation.

\subsection{Validation in dynamical spacetime: Symmetry-preserving test}
We validate our implementation of the new fourth-order Riemann solver in a dynamical spacetime by performing a symmetry-preserving test in a BNS merger NR simulation proposed in Ref.~\cite{Gao:2025nfj} (see the evolution code, grid setup, initial data, and EOS for details in the next section). 

To recap, we employ a specific compiler option {\tt -fp-model strict} for the Intel compiler to control a round-off error, and the symmetry-preserving technique (see Ref.~\cite{Gao:2025nfj} for details). Since this compiler option significantly reduces the computation speed, we employ a BNS configuration {\tt APR4\_14\_14} with the APR4~\cite{Akmal:1998cf} as an NS EOS, $m_1=m_2=1.4M_\odot$ where $m_{1,2}$ denotes the ADM mass of each NS in isolation, and the initial orbital separation $D_0=30M_\odot$ as an initial data (ID) in this specific test. The ID should possess a $\pi$-symmetry in machine precision. The grid setup is $N=41~(\Delta x=0.181M_\odot)$. 
We do not impose any spatial symmetry except for the orbital plane symmetry during the simulation. 

\begin{figure*}
    \centering
    \includegraphics[scale=0.34]{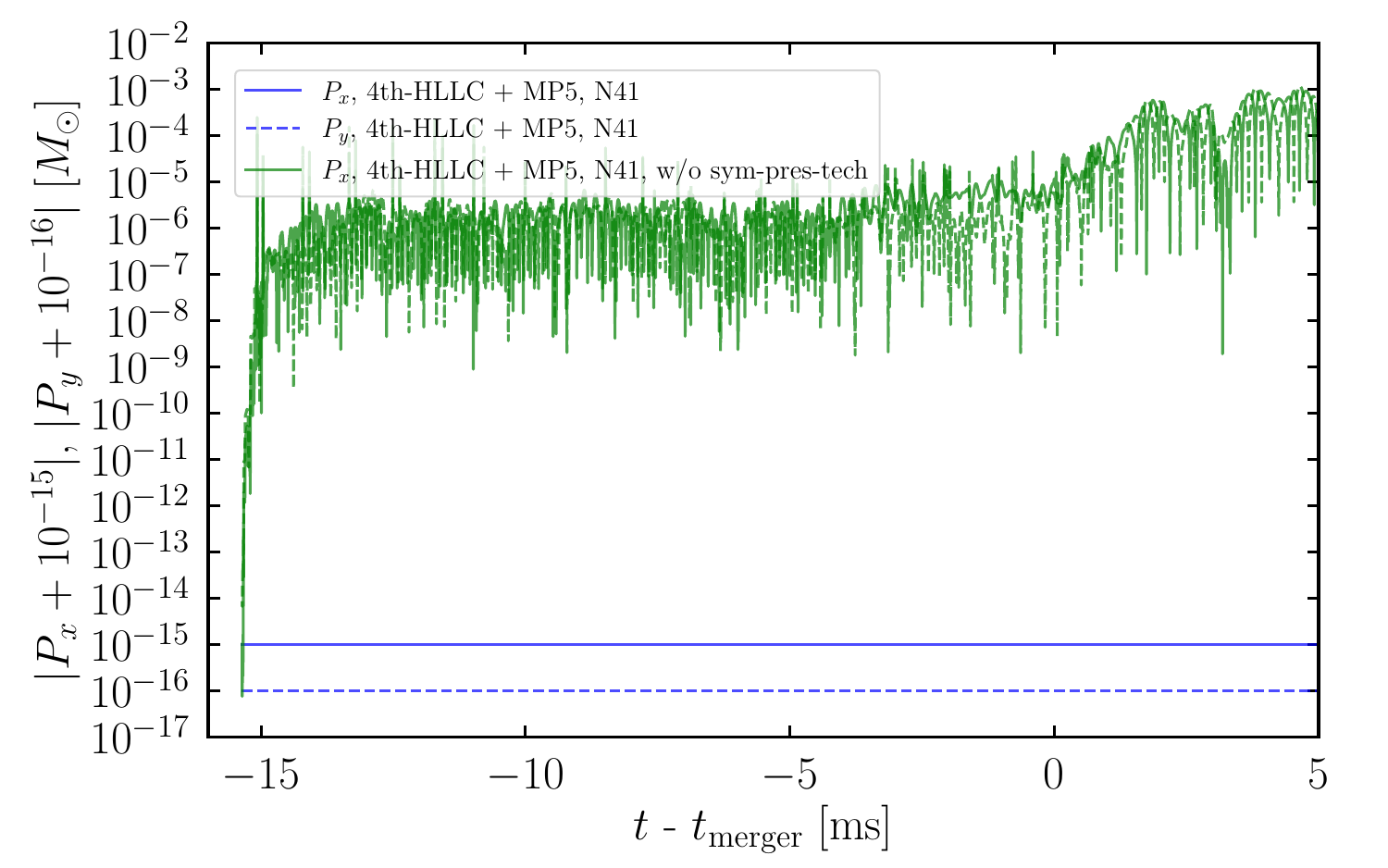}
    \includegraphics[scale=0.34]{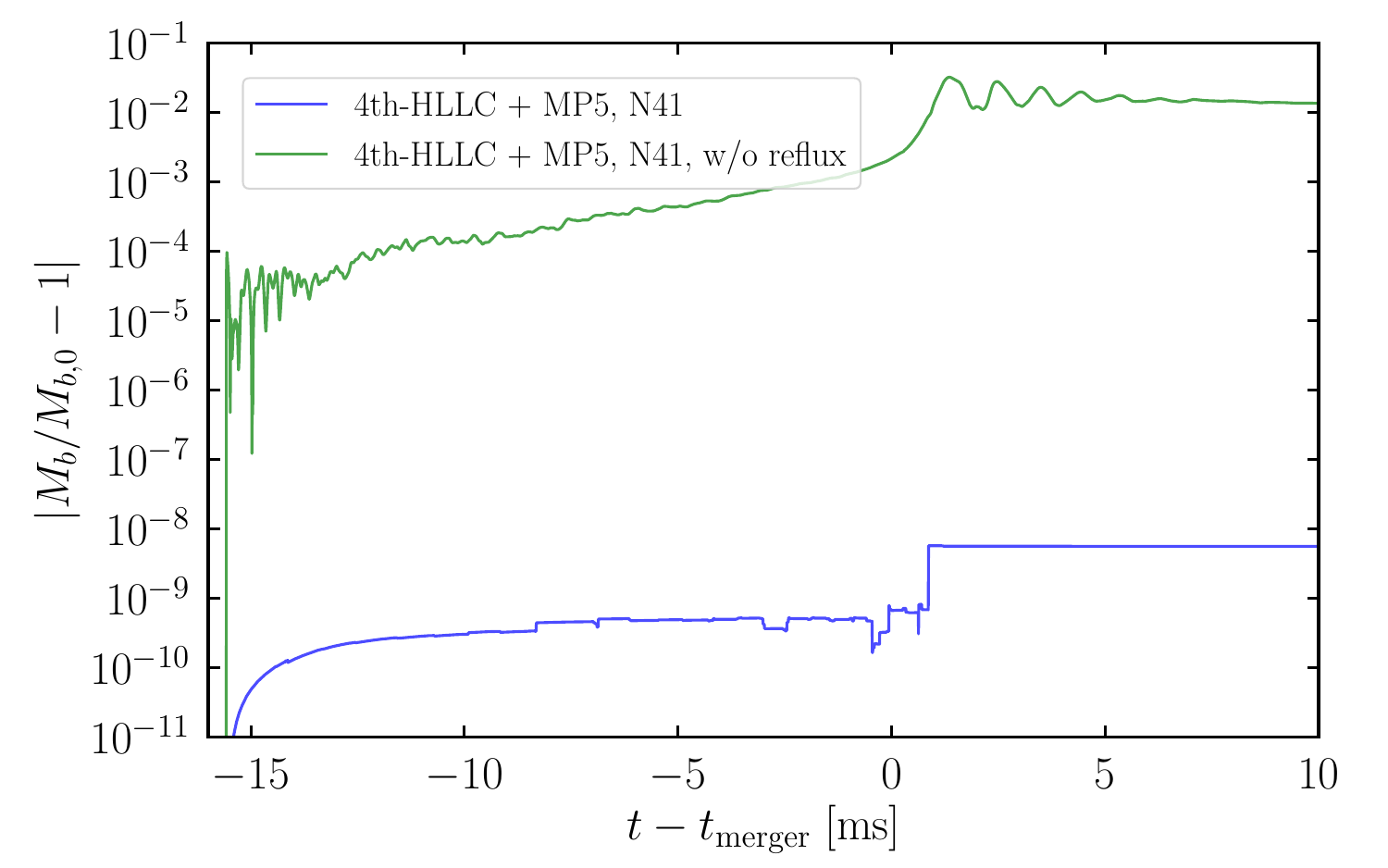}
    \caption{(Left) Conservation of the linear momentum in the x- and y-direction with blue lines in the symmetry-preserving test for the BNS merger ${\tt APR\_14\_14}$ with $N=41$ as a function of the post-merger time. To magnify the conservation accuracy, we add a small offset $10^{-15}$ and $10^{-16}$ to the x- and y-components, respectively. The green curves show the result of the symmetry-preserving test, where we turn off the symmetry-preserving technique while keeping the compiler option {\tt -fp-model strict}.
    (Right) Error of the baryon mass conservation as a function of the post-merger time, where $M_{b,0}$ denotes the initial baryon mass of the BNS (blue curve). The green curve is the result of the same model and grid setup, where we turn off the reflux prescription and conservative prolongation.
    }
    \label{fig:Sym-preserv}
\end{figure*}

The left panel of Fig.~\ref{fig:Sym-preserv} depicts the conservation of the linear momentum in the x- and y-direction in the symmetry-preserving test, where we define the linear momentum by
\begin{align}
P_{x/y} = \int \sqrt{\gamma} D h u_{x/y} {\mathrm d}^3x.
\end{align}
$\gamma$, $D$, $h$, and $u_{x/y}$ denote the determinant of the three spatial metric, conserved mass density, specific enthalpy, and x/y-component of the four velocity, respectively. The merger time $t_{\rm merger}$ is defined by the peak amplitude time of gravitational waves. 

On the one hand, the figure shows that the $\pi$-symmetry is perfectly preserved without imposing the $\pi$-symmetry throughout the entire stage of the evolution (see the blue lines). We note that Ref.~\cite{Gao:2025nfj} gives the proof of the linear momentum conservation in a discretized form if the $\pi$-symmetry is preserved. On the other hand, the figure also shows that, if we do not employ the symmetry-preserving technique as many existing NR codes may not, the linear momentum conservation is not preserved, i.e., the $\pi$-symmetry breaking, in the symmetry-preserving test (see the green curves in Fig.~\ref{fig:Sym-preserv}). Particularly, the violation exhibits exponential growth after the merger. 

Therefore, without demonstrating that an NR code passes this stringent symmetry-preserving test, we cannot reject the possibility that a code may suffer from bugs or artifacts that enhance asymmetry. We should emphasize that it is a critical code validation {\it before} exploring the putative $m=1$ instability in BNS mergers~\cite{Radice:2016,Nedora:2019jhl,East:2015vix,Lehner:2016wjg} since Ref.~\cite{Gao:2025nfj} shows that the violation of the linear momentum conservation highly correlates with the $m=1$ mode growth.

The right panel of Fig.~\ref{fig:Sym-preserv} plots the relative error of the baryon mass conservation. Although the spurious baryon mass increase happens in the inspiral phase due to the atmosphere, it stays $O(10^{-7})\%$ level throughout the entire evolution stage (see the blue curve). We note that, thanks to the reflux prescription as well as the conservative prolongation for the hydrodynamical variables, the baryon mass conservation is preserved fairly well even after the whole NS(s) is no longer covered by the finest AMR domains, i.e., a highly deformed hypermassive NS formed after the merger, whose size is much larger than the finest AMR domain(s). We also note that the reflux prescription without the conservative prolongation could result in $O(0.1)\%$ error in the baryon mass conservation~\cite{Dietrich:2015iva,Dietrich:2015pxa}. 

On the other hand, if we turn off the reflux prescription and conservative prolongation, an error of $O(1)$\% is unavoidable, particularly after the merger (see the green curve). It implies that, without quantifying the mass conservation error as many NR simulations did not (see Refs.~\cite{Kiuchi:2022,Kiuchi:2022nin,Kiuchi:2023obe,Gao:2025nfj,Haas:2016cop,Fields:2024pob,Cook:2023bag,Han:2025pho,Dietrich:2015iva,Dietrich:2015pxa} as an counter example which quantifies the mass conservation error), particularly, in the case that we do not employ the reflux prescription nor conservative prolongation, the ejecta mass smaller than $O(10^{-2})M_\odot$ could be masked by the conservation error. 
With these findings, we conclude that our code passes the symmetry-preserving test. 

\begin{figure*}
    \centering
    \includegraphics[scale=0.34]{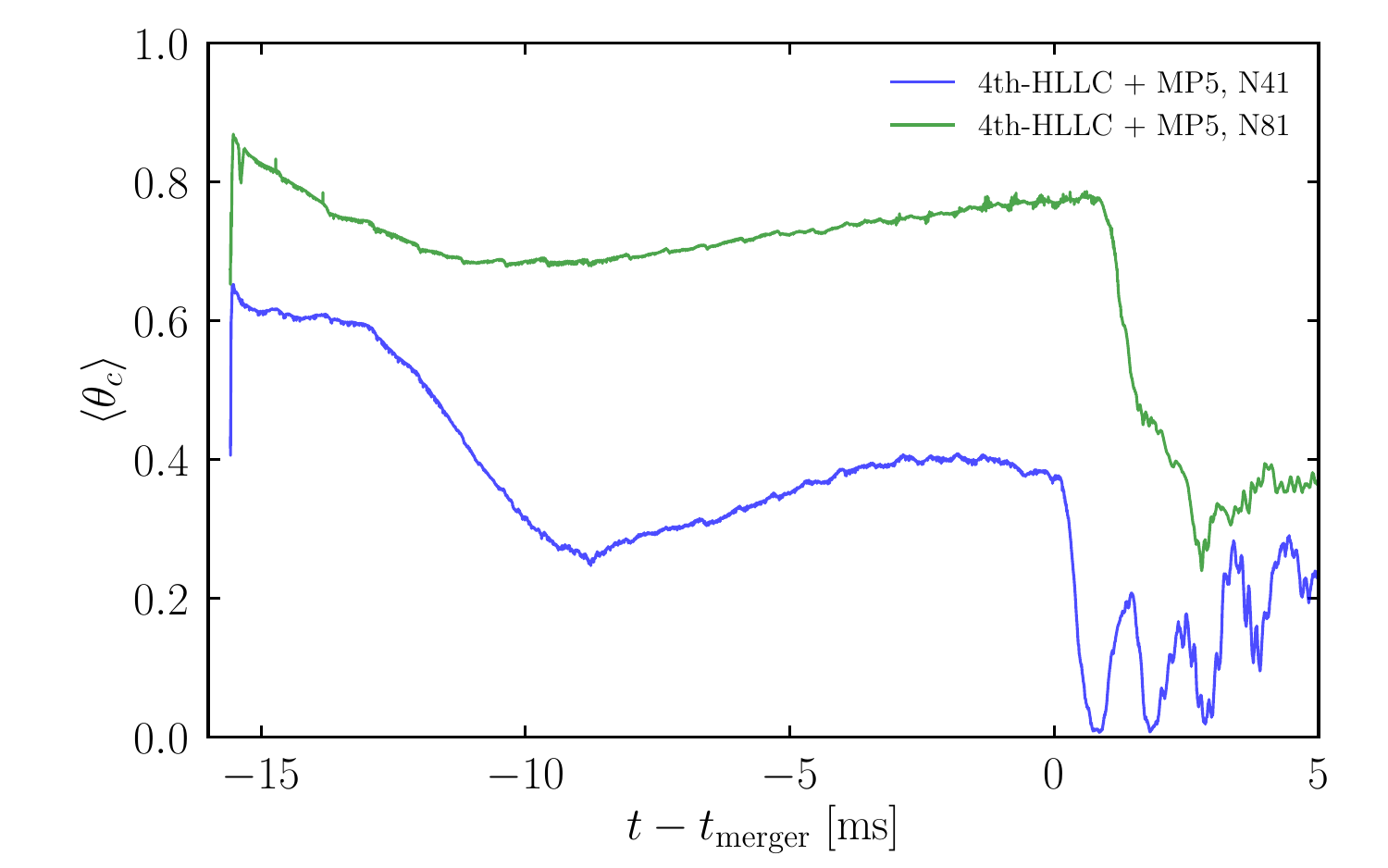}
    \includegraphics[scale=0.36]{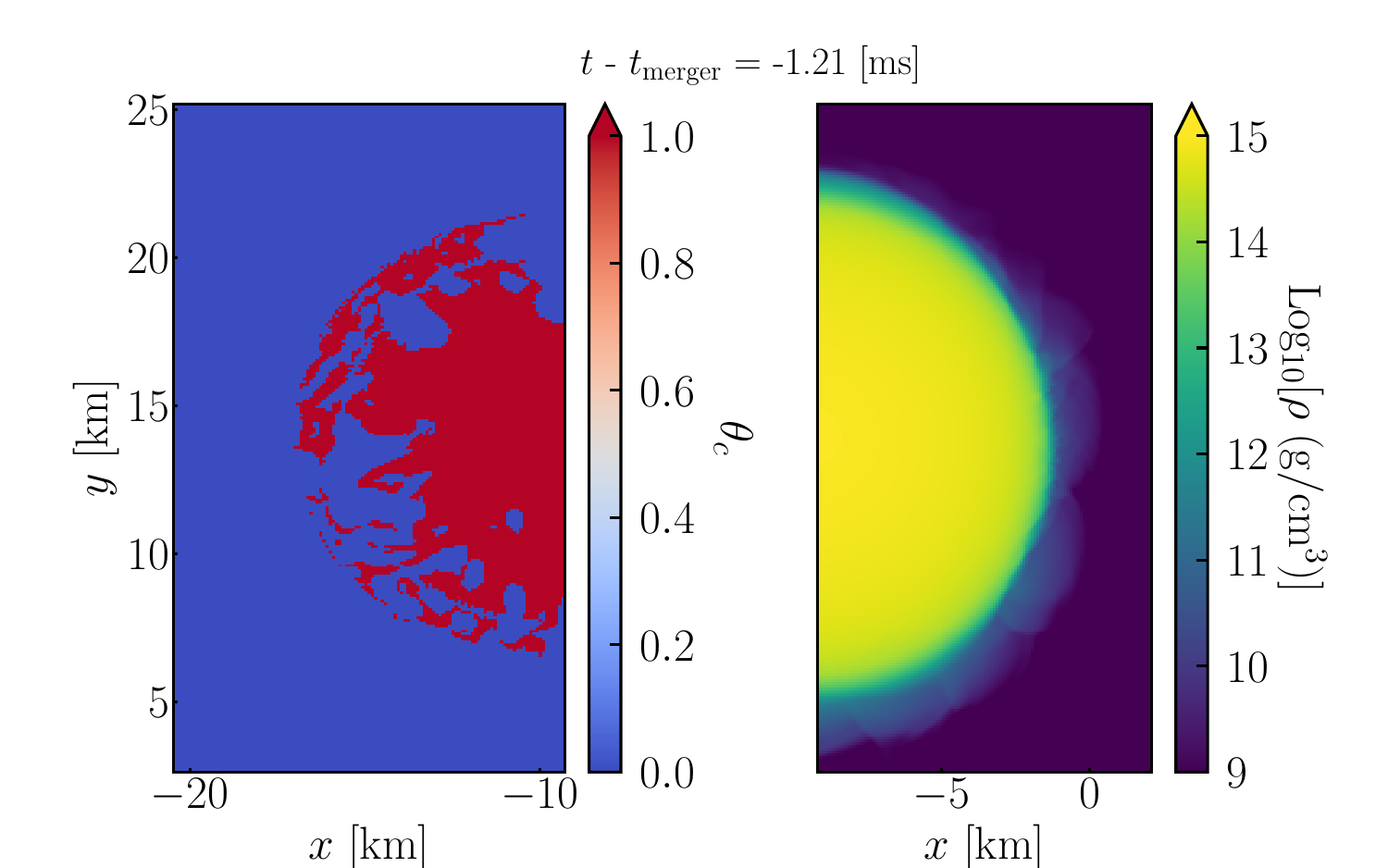}
    \caption{(Left) A density-weighted shock detector in the fourth-order HLLC Riemann solver as a function of the post-merger time. The BNS model is the same as in Fig.~\ref{fig:Sym-preserv} (blue curve). The green curve is the simulation with $N=81$. (Right) The rest-mass density and $\theta_c$ profile on the orbital plane at $t-t_{\rm merger}\approx -1$ ms for the simulation with $N=81$.
    }
    \label{fig:Sym-preserv2}
\end{figure*}

Finally, we quantify what fractions of the NSs are evolved with the fourth-order finite-volume Riemann solver by calculating a density-weighted shock detector: 
\begin{align}
\langle \theta_c \rangle \equiv \frac{\int \sqrt{\gamma}D \theta_c {\mathrm d}^3x}{\int \sqrt{\gamma}D {\mathrm d}^3x}.
\end{align}
The left panel of Fig.~\ref{fig:Sym-preserv2} shows that the fourth-order HLLC Riemann solver activates in a large fraction $\sim 60\%$ of the NSs during the very early inspiral phase, and the activation region drops to $\sim 20$\% at $t-t_{\rm merger}\approx -10$ ms. Then, it increases up to $\sim 40\%$ until $t-t_{\rm merger}\approx 0$ ms, and it drops again after the merger due to the shock generation during the hypermassive NS oscillations. 

We found that by improving the resolution, the fourth-order HLLC Riemann solver activation region drastically increases. In a higher resolution run with $N=81~(\Delta x=0.093M_\odot)$, the fourth-order Riemann solver activates in a $\gtrsim 70$\% region of the NSs during the inspiral phase and $\approx 30$--$40$\% in the early post-merger phase (see the green curve in the left panel and the right panel of Fig.~\ref{fig:Sym-preserv2}).
Therefore, we expect that the fourth-order Riemann solver will improve the accuracy of the inspiral and post-merger dynamics in BNS mergers. 

\section{Gravitational waves from a binary neutron star merger}\label{sec:result}
We quantify the performance of the new fourth-order accurate Riemann solver in an intermediate-term BNS merger simulation with high resolutions. 

\subsection{Binary neutron star model}
We employ a non-spinning equal-mass binary configuration with $m_1=m_2=1.35M_\odot$ as an ID of the simulation. The EOS for the neutron star matter is assumed to be the H4~\cite{Glendenning:1991es}. The initial orbital angular frequency is $m_0\Omega_0 \simeq 0.0161$ with $m_0=m_1+m_2$, which corresponds to $D_0=40M_\odot$ in the coordinate employed in an ID generator (see below). 
With this setup, we expect the BNS to experience an inspiral phase with $\approx 12$--$13$ orbits. We build a quasi-equilibrium BNS configuration by the public spectral library FUKA~\cite{Papenfort:2011}. We also employ an iterative orbital eccentricity reduction procedure to reduce a residual orbital eccentricity down to $O(10^{-4})$~\cite{Papenfort:2011}. 

\subsection{Evolution code and grid setup}\label{sec:NR_code}
We employ the adaptive mesh refinement (AMR) NR code {\tt SACRA-MPI}~\cite{Yamamoto:2008js,Kiuchi:2017zzg,Kiuchi:2019lls} to evolve the BNS in full general relativity. {\tt SACRA-MPI} employs the BSSN-puncture formulation~\cite{Shibata:1995,Baumgarte:1998te,Campanelli:2005dd} combined with the Z4c constraint propagation prescription~\cite{Hilditch:2012fp}. We discretize the geometrical variables with a fourth-order accurate finite difference scheme, and employ RK4 for the time integration. Concretely, we employ $\kappa_1=0.005M_\odot^{-1}$ and $\kappa_2=0$ for the Z4c prescription, and $\eta_B=0.03M_\odot^{-1}\sim 0.1/m_0$ for the damping parameter in the moving-puncture gauge~\cite{Hilditch:2012fp}. 
The fourth/second-order accurate HLLC Riemann solvers~\cite{Yamamoto:2008js,Kiuchi:2022} are employed with the fifth/third-order accurate cell reconstruction MP5/PPM~\cite{Suresh:1997}, respectively. The latter one is originally implemented in our code~\cite{Kiuchi:2022}. 

To cover a wide dynamical range of the problem, {\tt SACRA-MPI} employs the $2:1$ refinement for the box-in-box algorithm. To guarantee the baryon mass conservation, we employ a cell-centered grid structure with the reflux prescription. 
For all the geometrical variables, the prolongation/restriction is done with sixth-order/fourth-order accurate Lagrange interpolation, respectively. To respect the conservation law, {\tt SACRA-MPI} implements the conservative prolongation for the hydrodynamical variables. 

In simulations reported in this paper, {\tt SACRA-MPI} employs six non-moving AMR domains whose origin is fixed to the center of mass of the binary, and two sets of four moving AMR domains, which follow the NS orbital motion. Each domain has the same grid number $N$ and the x-coordinate in each domain is defined by $x_{({\rm lv})j} = x^o_{(\rm lv)} + ( j + 1/2 )\Delta x_{\rm lv},~j\in [-N-1,N]$ where $x^o_{(\rm lv)}$ denotes the $x$-coordinate origin of the AMR level of lv, i,e. ${\rm lv}=0,\cdots 9$ where $0~(9)$ denotes the coarsest (finest) AMR domain. The orbital plane symmetry is assumed. 

When we export the FUKA ID to {\tt SACRA-MPI}, we keep in mind that the distance between the coordinate origin of the two finest AMR domains, i.e., $x^o_{9}$ and its counterpart, should agree with the initial orbital separation we specify in FUKA in machine precision, irrespective of employed grid resolution. 
For example, if we directly specify the grid resolution, e.g., $\Delta x_9=0.1M_\odot$, in the FUKA exporter, a small truncation error causes a tiny, but non-negligible deviation from $D_0$ in the initial separation of the two finest AMR domains. Also, the deviation depends on the grid setup. It implies that the IDs with different grid resolutions may not be physically equivalent as long as we stick to this way to set up the grid resolution, even if we export the IDs from the exact same FUKA spectral data. It could be an obstacle for the convergence study in high-precision inspiral gravitational waveforms. At the same time, we should keep the finest AMR domain size $L$ in machine precision, irrespective of the employed grid resolutions. 

To achieve these requirements, we determine the grid size of the finest domain as follows. First, given the grid resolution $\Delta x_9$, which is yet specified, the domain size $L$ should be~\footnote{Since the cell center grid of the outermost cell in the x-direction is $x^o_9 + \left(N+1/2\right)\Delta x_9$, the AMR domain boundary should be $x^o_9 + \left(N+1\right)\Delta x_9$.} 
\begin{align}
L = 2\left(N+1 \right) \Delta x_9.
\end{align}
Then, we introduce an integer parameter $i$, which controls how many finest AMR domains fit between $x_9^o$ and its counterpart:
\begin{align}
i = \frac{D_0}{L} = \frac{D_0}{2(N+1)\Delta x_9}.
\end{align}
This equation determines the grid resolution $\Delta x_9$. However, we cannot specify $i$ freely since the size of the finest domain should be greater than the NS radius:
\begin{align}
L \ge 2 R_{\rm NS},
\end{align}
where $R_{\rm NS}$ is the coordinate radius of the NS in binary, and FUKA outputs it. In our model, $R_{\rm NS}\approx 7.4$--$7.6M_\odot$ depending on the positive or negative side from the NS center. Namely, in the current setup, $i=D_0/L \le D_0/(2R_{\rm NS})\approx 2.63$. 
This way to specify the grid resolution guarantees that the separation between $x_9^o$ and its counterpart agrees with $D_0$ in machine precision, irrespective of the employed grid resolution. We choose $i=2$ in our grid setup, i.e., $L=20M_\odot$. To check the convergence in simulations, we vary $N=189,157,125,109$ in this paper. Table~\ref{tab:model} summarizes the BNS model and grid setup.

During the evolution, we employ a hybrid EOS with the piece-wise-polytrope~\cite{Read:2008iy} and $\Gamma$-law:
\begin{align}
P = P_\mathrm{c}(\rho) + \left(\Gamma_{\rm th}-1\right) \rho \left(\epsilon-\epsilon_{\rm c}(\rho)\right),
\end{align}
where $P,P_{\rm c},\rho,\epsilon,$ and $\epsilon_{\rm c}$ denote the pressure, the cold part of the pressure, rest-mass density, specific internal energy, and the cold part of the specific internal energy, respectively. $P_{\rm c}$ and $\epsilon_{\rm c}$ should obey the first-law of the thermodynamics:
\begin{align}
{\mathrm d}\epsilon_{\rm c} = -P_{\rm c} {\mathrm d}\left(\frac{1}{\rho}\right).
\end{align}
We employ a four-segments piece-wise-polytrope representation of the H4 EOS for $P_c(\rho)$ and $\epsilon_c(\rho)$~\cite{Read:2008iy}. 
$\Gamma_{\rm th}$ describes an efficiency of the shock heating and we take $5/3$ throughout the simulations.

\begin{table*}
\caption{Model, EOS, ADM mass of each NS in isolation, tidal deformability of $1.35M_\odot$ NS, initial orbital coordinate separation, initial orbital frequency, the size of the finest domain, and grid setup. ${\rm lv_{nmv}/lv_{mv}}$ denotes the non-moving/moving domains, respectively, where ${\rm lv}=0$ is the coarsest level. 
$\Delta x_{9}$ is the grid resolution of the finest AMR domains for $N=189,157,125,109$. 
The last column denotes the residual orbital eccentricity. 
}\label{tab:model}
\begin{tabular}{ccccccccccc}
\hline\hline
Model & EOS & $m_1/m_2$ $[M_\odot]$ & $\Lambda_{1.35}$ & $D_0$ $[M_\odot]$ & $m_0\Omega_0$ & $L$ $[M_\odot]$ & ${\rm lv}_{\rm nmv}$ & $\rm lv_{mv}$ & $\Delta x_{9}$ $[M_\odot]$ & $e_0$\\
\hline\hline
${\tt H4\_135\_135\_00161}$ & H4 & 1.35/1.35 & 1157 & 40 & 0.0161 & 20 & 0--5 & 6--9 & $[0.0526,0.0633,0.0794,0.0909]$ & $1.0\times10^{-4}$\\
\hline
\end{tabular}
\end{table*}

\subsection{Performance of the higher-order Riemann solver}
\subsubsection{Trajectory}
\begin{figure*}
    \centering
    \includegraphics[scale=0.5]{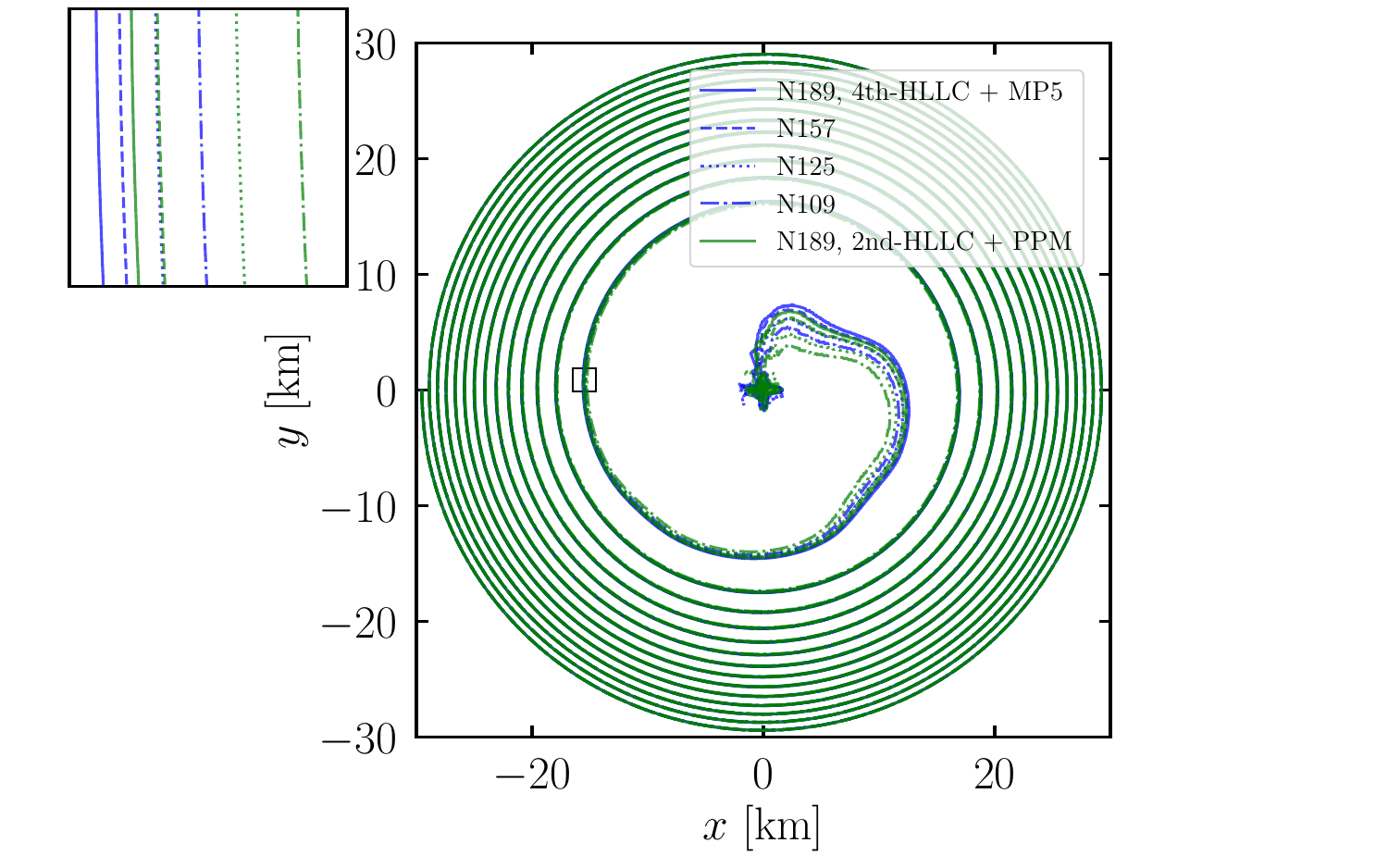}
    \caption{A trajectory of the center of mass of the NS1 during the BNS merger {\tt H4\_135\_135\_00161}. For visibility, the trajectory of the NS2 is not shown. The blue (green) curves denote the fourth-order (second-order) HLLC Riemann solver. The solid, dashed, dotted, and dashed-dotted curves are for $N=189, 157, 125$, and $109$, respectively. The inset is a close-up of the trajectory in a region shown with a small box in the main plot, and its scale is $0.5$~km. We note that the blue-dotted curve and the green-dashed curve almost overlap in the inset.
    }
    \label{fig:trj}
\end{figure*}

Figure~\ref{fig:trj} plots a trajectory of the center of mass of the NS1. The center of the mass in the x-direction is defined by
\begin{align}
x_{\rm NS1, COM} \equiv \frac{\int_{\rm NS1} \sqrt{\gamma}D x {\it d}^3x}{M_{b,{\rm NS1}}},
\end{align}
where $M_{b,{\rm NS1}}$ is the baryon mass of the NS1, and the integration is done for the NS1 region. 
The center of mass in the y-direction is defined in a similar way. The trajectories with the different Riemann solvers and different resolutions are indistinguishable on this scale, except for the final moment of the merger. 

If we look at the close-up of the trajectory shown in the inset, we have two observables. One is that, given the Riemann solver, the orbital decay due to the numerical dissipation is slower in the higher-resolution case (see the solid, dashed, dotted, and dashed-dotted curves). The second point is that, interestingly, the orbital decay with the fourth-order HLLC with $N=157,125$, and $109$ (the blue-dashed, dotted, and dashed-dotted curve) is slower than the second-order HLLC with $N=189,157$, and $125$ (the green-solid, -dashed, and -dotted curves), respectively. 
It indicates that the suppression of the numerical dissipation with the fourth-order HLLC is significant. 

Since the orbital decay due to the numerical dissipation causes a phase error in gravitational waves, this improvement indicates that the fourth-order HLLC Riemann solver would be a key ingredient to derive high-precision gravitational waveforms in BNS merger NR simulations. 
As described in Sec.~\ref {sec:implementation}, the fourth-order Riemann solver requires the additional primitive recovery, and we measured a $\sim 16\%$ increase in the CPU times compared to the second-order Riemann solver with the same grid setup. However, this drawback is compensated by the less dissipative feature of the fourth-order Riemann solver. For example, the simulation cost with $N=157,125,109$ is $\approx 52,60,40\%$ cheaper than that with $N=189,157,125$, respectively. Because the fourth-order HLLC with $N=157,125,109$ is less dissipative than (or at least comparable to) the second-order HLLC with $N=189,157,125$, we gain $\sim 38,44,24\%$ net CPU-time decrease, respectively, to derive the same (or maybe better) quality inspiral gravitational waveforms. We furthermore quantify the performance of the new fourth-order accurate Riemann solver below.

\subsubsection{Gravitational waves}
\begin{figure*}
    \centering
    \includegraphics[scale=0.45]{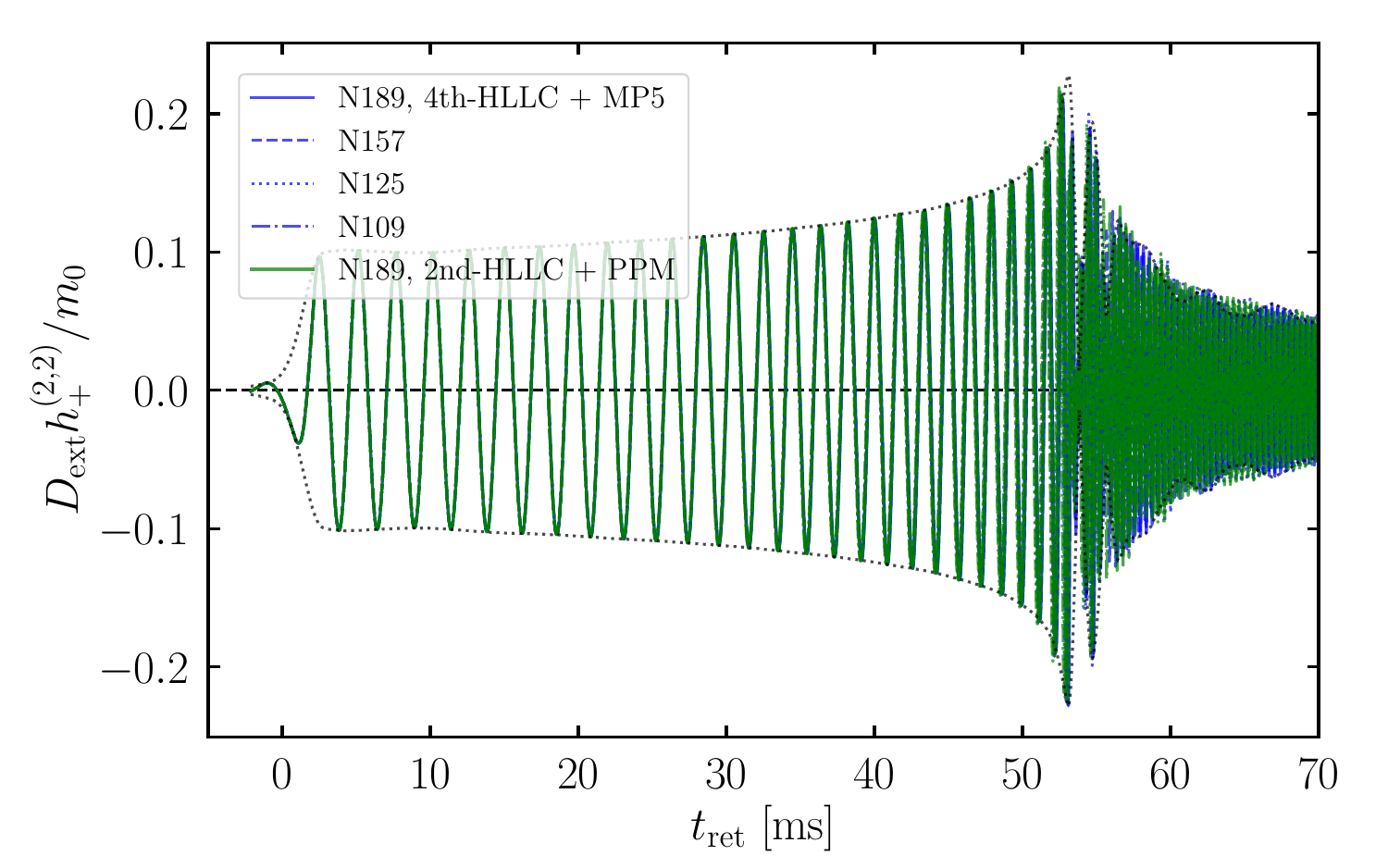}\\
    \includegraphics[scale=0.45]{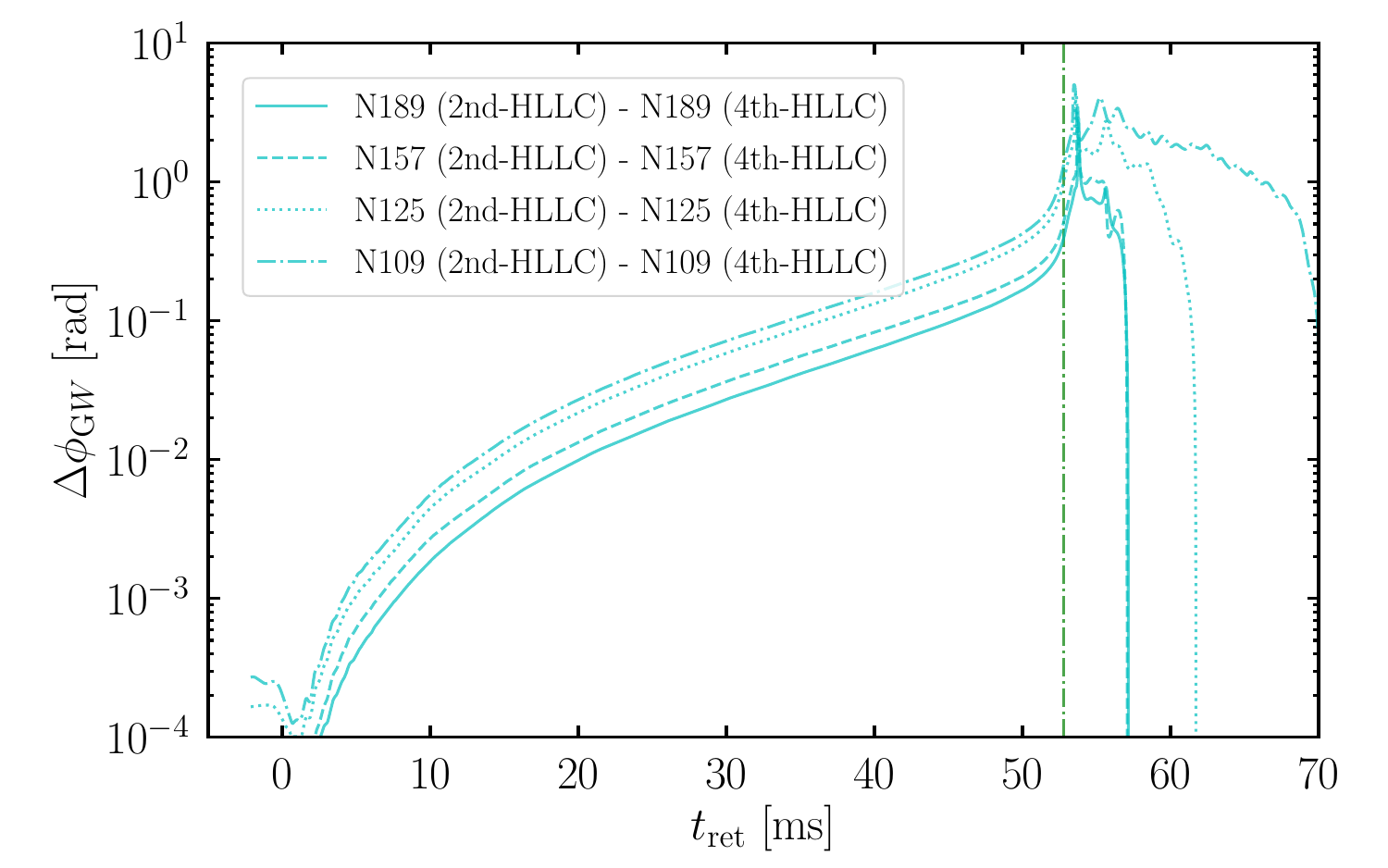}\\
    \includegraphics[scale=0.45]{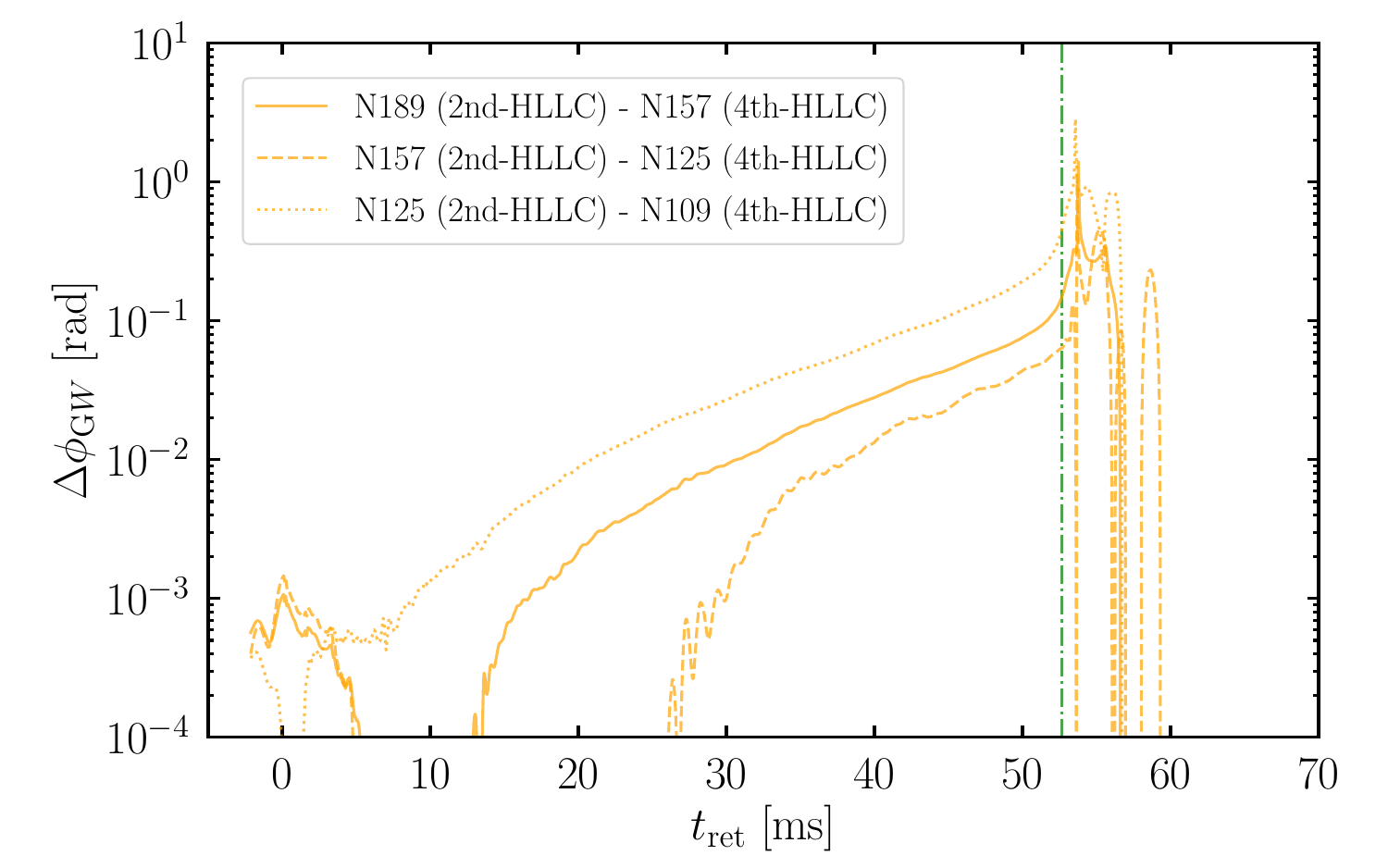}
    \caption{(Top) The $+$-mode of the gravitational waves with $(l,m)=(2,2)$ as a function of the reterted time. The blue (green) curve shows the fourth (second)-order HLLC Riemann solver with MP5 (PPM) cell reconstruction. The solid, dashed, dotted, and dashed-dotted line styles denote $N=189,157,125$, and $109$, respectively. The black-dotted curves are the envelopes of the gravitational waves.  (Middle) Phase difference between the fourth-order and second-order HLLC with $N_1=N_2=189,157,125$, and $109$ (cyan-solid,-dashed,-dotted, and -dashed-dotted curve), respectively. (Bottom) The orange-solid, -dashed, and -dotted curves are the phase difference between the second-order HLLC and the fourth-order HLLC with $(N_1,N_2)=(189,157),(157,125)$, and $(125,109)$, respectively. The vertical dashed-dotted line presents the merger time for the second-order HLLC with $N=109$ in the middle and bottom panels. 
    }
    \label{fig:GW}
\end{figure*}

Figure~\ref{fig:GW} plots the $+$-mode of the gravitational waves for $(l,m)=(2,2)$ as a function of the retarded time. We employ the fixed-frequency method~\cite{Reisswig:2010di} to calculate $h_{+,\times}$ from $\psi_4$ extracted on the sphere with the extraction radius $D_{\rm ext}=400M_\odot$ during the simulation. We also extrapolate the gravitational waves to infinity by Nakano's method~\cite{Nakano:2015rda} (see Ref.~\cite{Kiuchi:2019kzt} for details). 

The middle and bottom panels in Fig.~\ref{fig:GW} plot the gravitational wave phase difference between two resolutions, $N_1$ and $N_2$, and two Riemann solvers (RS), ${\rm RS}_1$ and ${\rm RS}_2$, defined by
\begin{align}
\Delta \phi_{\rm GW} = \phi_{{\rm GW},N_1}^{\text{RS}_1} - \phi_{{\rm GW},N_2}^{\text{RS}_2}
\end{align}
as a function of the retarded time where the phase $\phi_{\rm GW}$ is calculated by
\begin{align}
h_+(t)-ih_\times(t) = A(t) {\rm e}^{-i\phi_{\rm GW}(t)}.
\end{align}
The middle panel shows the phase difference between the second-order HLLC as $\rm RS_1$ and the fourth-order HLLC as $\rm RS_2$ with $N_1=N_2=189,157,125$, and $109$ (cyan-solid, -dashed, -dotted, and -dashed-dotted curve), respectively. The phase difference at the merger~\footnote{Throughout this section, the merger time is defined by the simulation with the second-order HLLC with $N=109$.} is $\approx 0.36,0.46,0.90$, $1.17$ rad out of the total phase $\approx 176$ rad for $N=189,157,125,$ and $109$, respectively. It demonstrates that the fourth-order HLLC is enable to produce higher-quality gravitational waveforms than the second-order HLLC, irrespective of the employed grid resolutions. 

The bottom plot shows the phase difference between the second-order HLLC as $\rm RS_1$ and the fourth-order HLLC as $\rm RS_2$ with $(N_1,N_2)=(189,157),(157,125),$ and $(125,109)$ (orange-solid, -dashed, and -dotted curves, respectively). The phase difference at the merger is $\approx 0.15,0.06,$ and $0.46$ rad. 
The positive phase difference at the merger demonstrates that the fourth-order HLLC can derive more accurate gravitational waveforms with a cheaper computational cost than the second-order HLLC. 

Finally, we quantify the convergence order and the residual phase error of the gravitational waves derived in this paper. We begin with the standard ansatz for the gravitational waves phase to estimate the convergence order and the continuum limit:
\begin{align}
\phi_{\rm GW}(t;N) = \phi_{\rm GW}^\infty(t) + a(t) \left(\frac{\Delta x_N}{\Delta x_{189}}\right)^{p_\mathrm{conv}(t)}, \label{eq:ansatz1}
\end{align}
where $\Delta x_N(\Delta x_{189})$ denotes the grid resolution with the grid number $N(N=189)$. Normally, we estimate the three unknowns $\phi^\infty_{\rm GW},a$, and $p_{\rm conv}$ by utilizing the simulation data $\phi_{\rm GW}(t;N)$ with (at least) three different $N$. As discussed in Ref.~\cite{Kuan:2025bzu}, the least square fitting for three unknowns with four (or more) resolutions could be plausible to determine the convergence order in a cleaner way. Therefore, we utilize the simulation data with all the resolutions to fit the three unknowns (see Appendix~\ref{appdx:sfcf} for the self-convergence study of the phase error). 

\begin{figure*}
    \centering
    \includegraphics[scale=0.45]{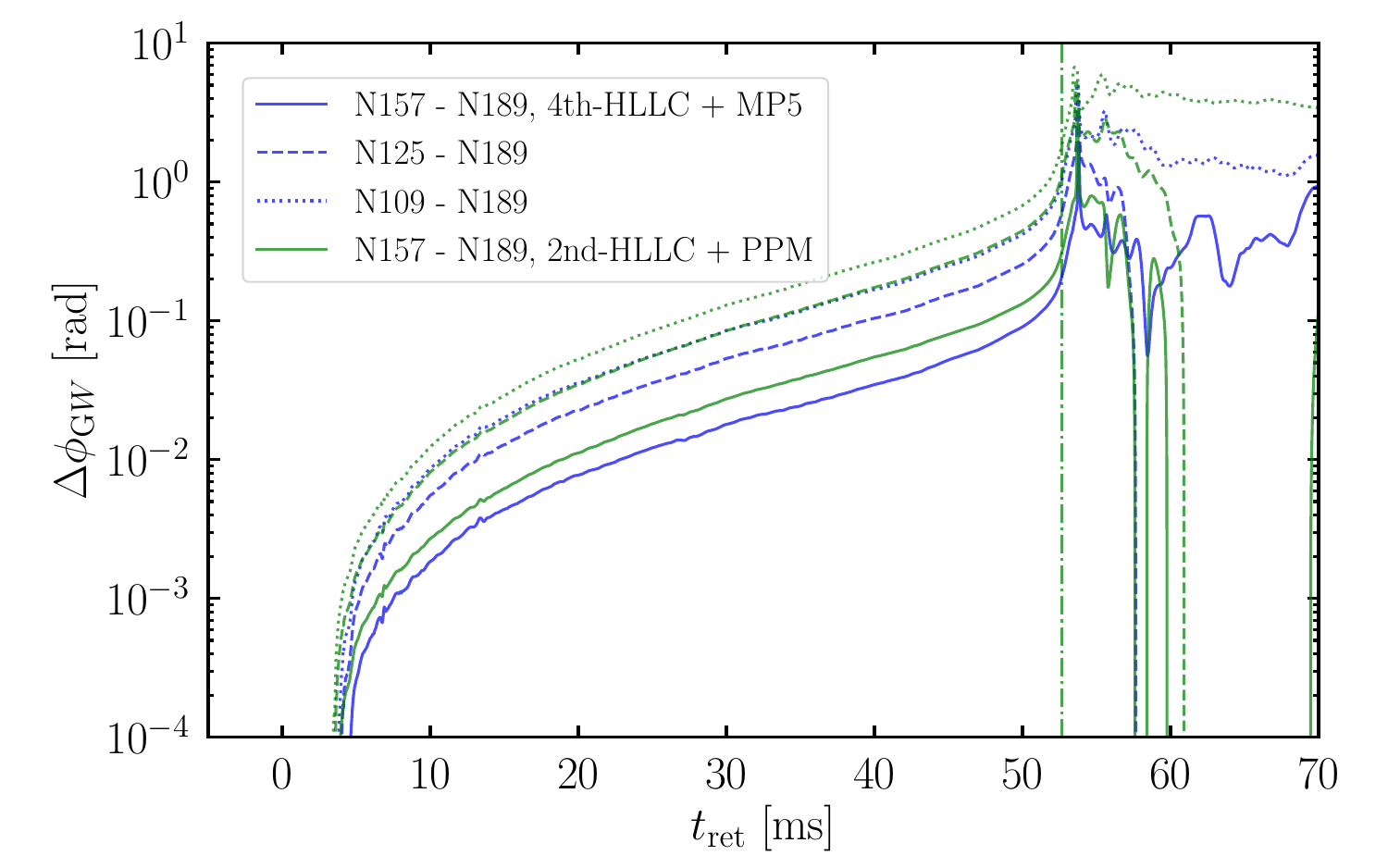}\\
    \includegraphics[scale=0.45]{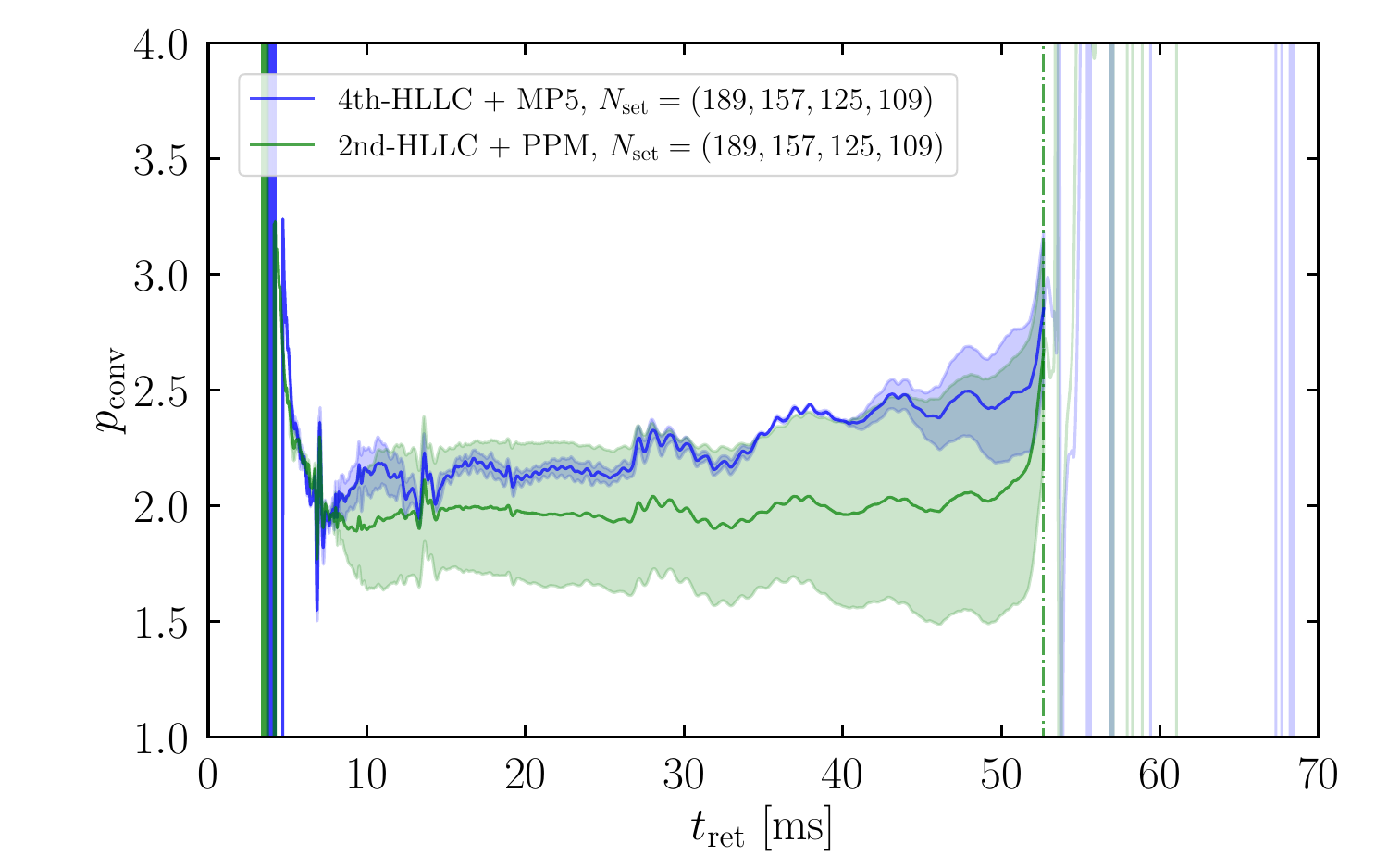}\\
    \includegraphics[scale=0.45]{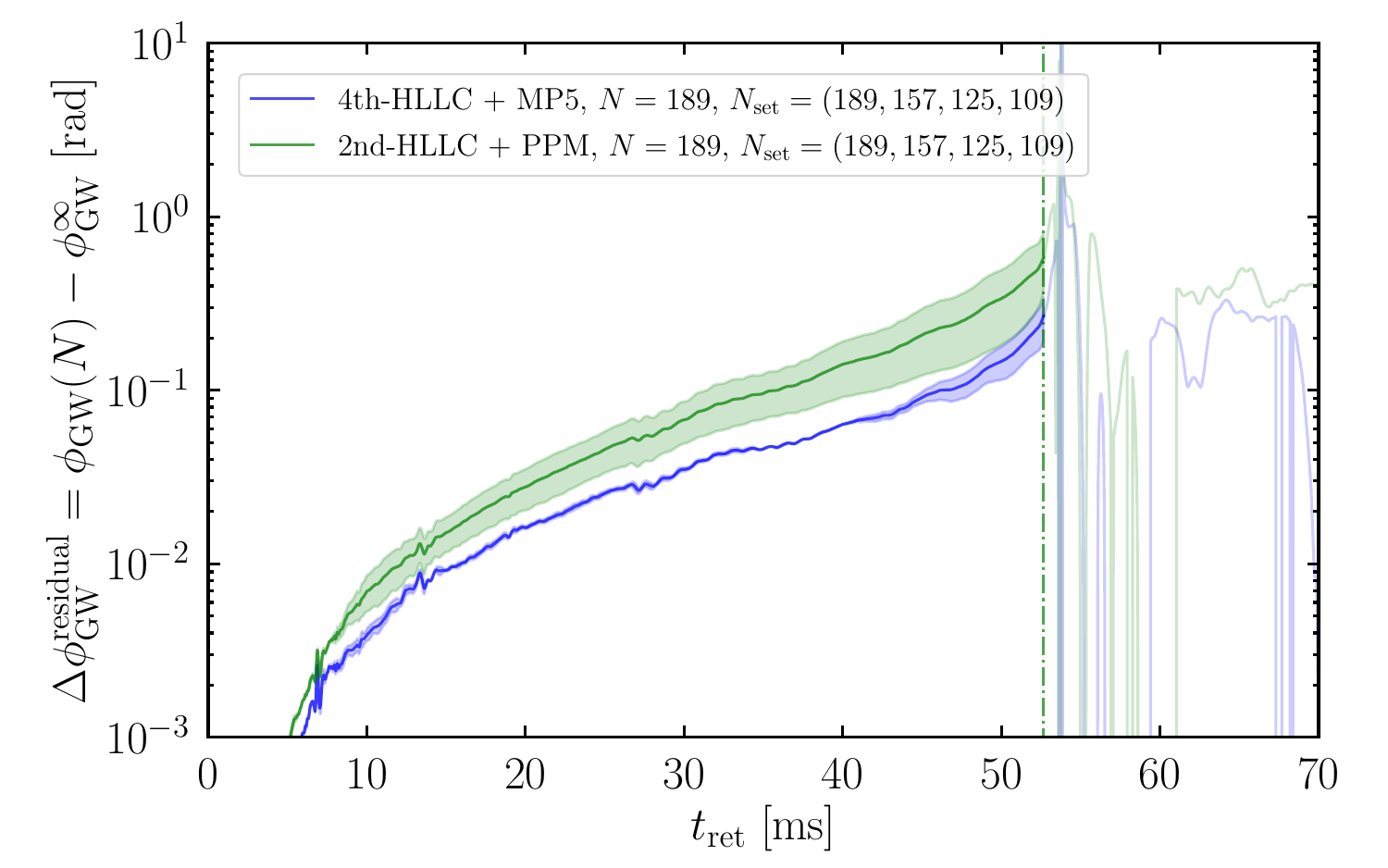}
    \caption{(Top) Gravitational wave phase error with the fourth-order HLLC (blue) and the second-order HLLC (green). 
    The solid, dashed, and dotted curves are for $(N_1,N_2)=(157,189), (125,189)$ and $(109,189)$, respectively. (Middle) The convergence order for the fourth-order HLLC (blue) and the second-order HLLC (green). 
    (Bottom) The residual phase error of the gravitational waves for the fourth-order HLLC (blue) and the second-order HLLC (green). 
    The vertical dashed-dotted line presents the merger time for the second-order HLLC with $N=109$ in all the panels. 
    In the middle and bottom panels, the color-shaded region denotes $1\sigma$ region in the fitting, and we make the curve semi-transparent after the merger for visibility.
    }
    \label{fig:GW2}
\end{figure*}

The top and middle panels of Fig.~\ref{fig:GW2} plot the gravitational wave phase error and the convergence order as a function of the retarded time. The color-shaded region is $1\sigma$ region in the fitting. 
On the one hand, the convergence order evolves from $\approx 2.1\pm 0.05 $ to $2.4\pm 0.27$ for $10~{\rm ms}\lesssim t_{\rm ret} \lesssim 51~{\rm ms}$ for the fourth-order HLLC~\footnote{The convergence order in the early phase $t_{\rm ret}\lesssim 10~{\rm ms}$ could be unreliable because of the junk radiation and residual orbital eccentricity (see the modulation in the envelope in the early phase in the top panel of Fig.~\ref{fig:GW}).}. 
On the other hand, $p_{\rm conv} \approx 2.0\pm 0.5$ for the second-order HLLC~\footnote{$p_{\rm conv}>2$ for the second-order HLLC could be an artifact due to the fitting because $p_{\rm conv}$ should be determined by the lowest-order accurate tools in our NR code, i.e., the second-order Riemann solver.}. The uncertainty in the convergence order with the fourth-order HLLC is significantly smaller than that with the second-order HLLC. 

The bottom panel of Fig.~\ref{fig:GW2} plots the residual phase error of the gravitational waves toward the continuum limit defined by 
\begin{align}
\Delta \phi_{\rm GW}^{\rm residual}=\phi_{\rm GW}(t;N)-\phi_{\rm GW}^\infty(t). \label{eq:residual}
\end{align} 
The residual error with $N=189$ at the merger is $\approx 0.27\pm 0.07$ rad for the fourth-order HLLC, and $\approx 0.58\pm 0.22$ rad for the second-order HLLC, i.e., a factor of $\approx 2$ improvement. 

\begin{figure*}
    \centering
    \includegraphics[scale=0.35]{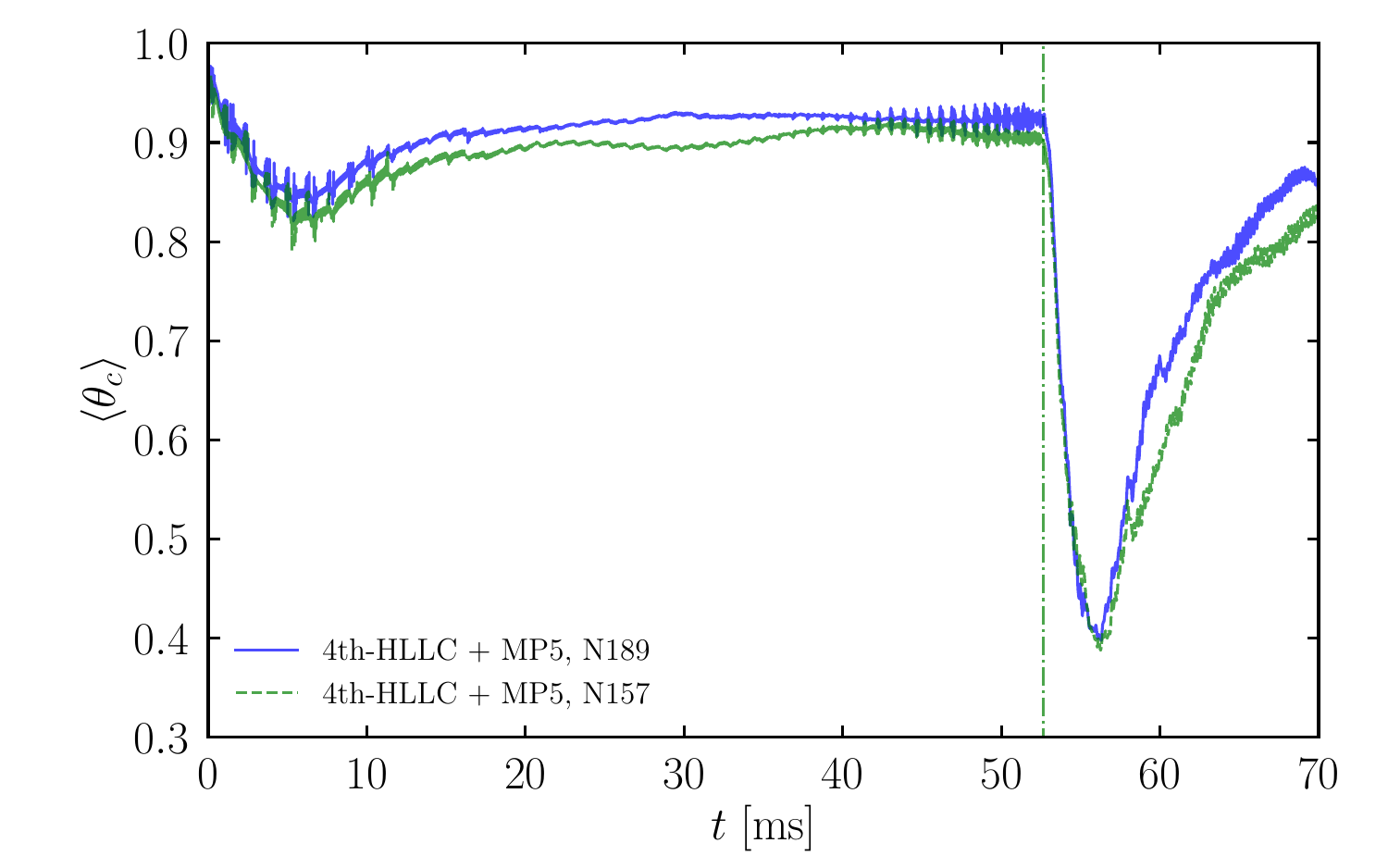}
    \includegraphics[scale=0.35]{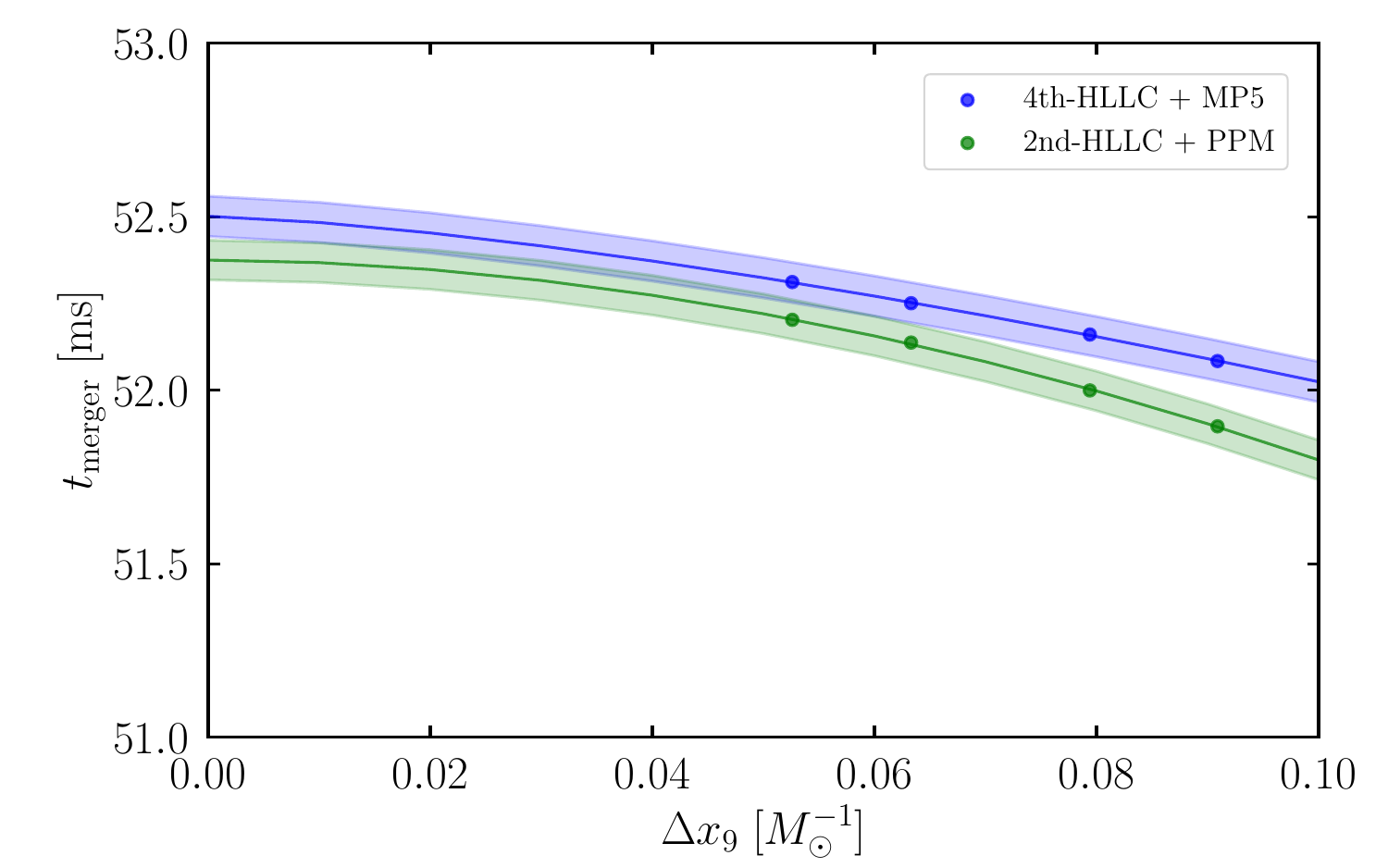}
    \caption{(Left) The density-weighted shock detector as a function of time. The blue and green curves present with $N=189$ and $157$, respectively. The vertical dashed-dotted curve is the merger time for the second-order HLLC with $N=109$. (Right) The merger time as a function of the grid resolution. The blue (green)-solid curve is the fitting curve with $1\sigma$ region for the fourth (second)-order HLLC.
    }
    \label{fig:theta_c_2}
\end{figure*}

\section{Discussion}\label{sec:discussion}
In Eq.~(\ref{eq:ansatz1}), we assume $\phi_{\rm GW}^\infty,a,$ and $p_{\rm conv}$ do not depend on the grid resolution. There should be a caveat for this assumption. The convergence order $p_{\rm conv}$ is determined by the lowest-order accurate tools in our NR code, i.e., the Riemann solver (see Sec.~\ref{sec:NR_code} for the accuracy of each tool in detail). 
If we employ the second-order HLLC, we expect $1\le p_{\rm conv} \le 2$ since the Riemann solver will be reduced to the first-order in the presence of the shock. 
What fraction of the NSs is evolved with the second-order accuracy is automatically controlled by the flux-limiter function (the minmod function in our case~\cite{Kiuchi:2022}). This fraction may depend on the employed grid resolution, i.e., the finer the grid resolution is, the larger the fraction is. 

Similarly, if we employ the fourth-order accurate HLLC, we expect $1 \le p_{\rm conv} \le  4$. The shock detector $\theta_c$ and the flux-limiter function (MP5 in our case) control what fraction of the NSs is evolved with fourth-order accuracy. Again, this fraction may depend on the employed grid resolution. The left panel of Fig.~\ref{fig:theta_c_2} plots the density-weighted shock detector with $N=189$ and $157$ as a function of time. It supports our hypothesis, i.e., the finer the grid resolution is, the larger $\langle \theta_c \rangle$ is. 

Since the overall convergence order $p_{\rm conv}$ is determined by a mixture of the several accuracies (the second- and first-order for the second-order accurate Riemann solver, and the fourth-, second-, and first-order for the fourth-order accurate Riemann solver), the ansatz~(\ref{eq:ansatz1}) could have to be modified:
\begin{align}
\phi_{\rm GW}(t;N) = \phi_{\rm GW}^\infty(t;N_{\rm set}) + a(t;N_{\rm set}) \left(\frac{\Delta x_N}{\Delta x_{189}}\right)^{p_\mathrm{conv}(t;N_{\rm set})}, \label{eq:ansatz2}
\end{align}
where $N_{\rm set}$ denotes a set of the employed grid resolutions to determine the three unknowns $\phi_{\rm GW}^\infty,a$, and $p_{\rm conv}$. Therefore, we should check the convergence order, and the residual phase error would be improved in a finer resolution simulation in future work. 

We also have a caveat for the definition of the merger time, which is defined by the second-order HLLC with the lowest resolution in this paper. Because the merger time is improved by improving the resolution, the residual phase error at the merger~(\ref{eq:residual}) in this paper could underestimate the residual phase error in the continuum limit. The right panel of Fig.~\ref{fig:theta_c_2} plots the merger time as a function of the grid resolution and its continuum limit. The merger time estimated by the fourth-order and second-order HLLC is consistent within the uncertainty. Even in the highest resolution run, the merger time is underestimated by $\approx 0.17$--$0.18$ ms. A finer resolution simulation will validate the extrapolation to the continuum limit. 

The convergence property for the post-merger gravitational waveform is completely lost, particularly with the low resolution and/or the second-order HLLC (see the sign flip in the phase error in the top panel of Fig.~\ref{fig:GW2}). This behavior is observed in many other NR simulations for BNS mergers (see Ref.~\cite{Dietrich:2019kaq} for example). However, we speculate that our new fourth-order HLLC may mitigate this pathological behavior in the post-merger gravitational waves if we employ a super-high resolution with $N \ge 189$ as indicated by (i) the increase of the density-weighted shock detector by improving the resolution during the post-merger phase in the left panel of Fig.~\ref{fig:theta_c_2} and (ii) no sign flip in the phase error between $N=189$ and $157$ during the post-merger phase (at least until the end of the simulation) in the top panel of Fig.~\ref{fig:GW2}.

All the waveform data reported in this paper can be downloaded from the SACRA Gravitational Waveform Data Bank: \url{https://www2.yukawa.kyoto-u.ac.jp/~nr_kyoto/SACRA_PUB/catalog.html}. 
The model name is {\tt H4\_135\_135\_00161\_N\_RS} where $N=189,157,125,109$ and ${\rm RS=HLLC4/HLLC}$ for the fourth/second-order accurate HLLC Riemann solver. 
\section{Conclusion}\label{sec:conclusion}
We implement the fourth-order accurate finite-volume HLLC Riemann solver in the NR code {\tt SACRA-MPI}, and validate our implementation both in the flat and dynamical spacetime. 

By conducting the in-depth resolution study, we quantify the performance of the newly implemented fourth-order accurate finite-volume HLLC Riemann solver in the non-spinning equal mass BNS merger simulations. We found the improvement of the convergence order in the inspiral gravitational wave phase, $p_{\rm conv} \approx 2.1$--$2.4$, compared to the convergence order, $p_{\rm conv}\approx 2.0$, derived by the originally implemented second-order accurate finite-volume HLLC Riemann solver. We also found the improvement of the residual phase error in the inspiral gravitational waves at the merger by a factor of $\approx 2$ compared to the second-order HLLC Riemann solver. Furthermore, we demonstrate that the fourth-order HLLC Riemann solver is able to derive higher-quality gravitational waveforms with cheaper computational costs compared to the second-order HLLC Riemann solver. 

With these findings, we conclude that the fourth-order accurate finite-volume Riemann solver will be a key factor in deriving high-precision inspiral gravitational waveforms from BNS mergers. We should note that the higher-order finite-volume Riemann solver beyond the fourth order is straightforward thanks to the smart and simple strategy with the volume-averaged and point-wise variables invented in Ref.~\cite{Berta:2023zmr} (see Eq.~(\ref{eq:FVtoPW})). We plan to derive long-term inspiral gravitational waveforms from BNS mergers as a future project. 

\acknowledgments 

KK thanks the Computational Relativistic Astrophysics division members in AEI, especially Masaru Shibata, Kota Hayashi, and Hao-Jui Kuan. Numerical computations were performed on the clusters Sakura, Cobra, Raven, and Viper at the Max Planck Computing and Data Facility, also on FUGAKU in RCCS (hp25006). KK is in part supported by Grant-in-Aid for Scientific Research (grant No.~23K25869) of Japanese MEXT/JSPS.

\section*{Data availability} 
The data that support the findings of this article are openly available~\cite{SACRA_URL}.


\appendix

\section{Self-convergence test}\label{appdx:sfcf}
\begin{figure*}
    \centering
    \includegraphics[scale=0.35]{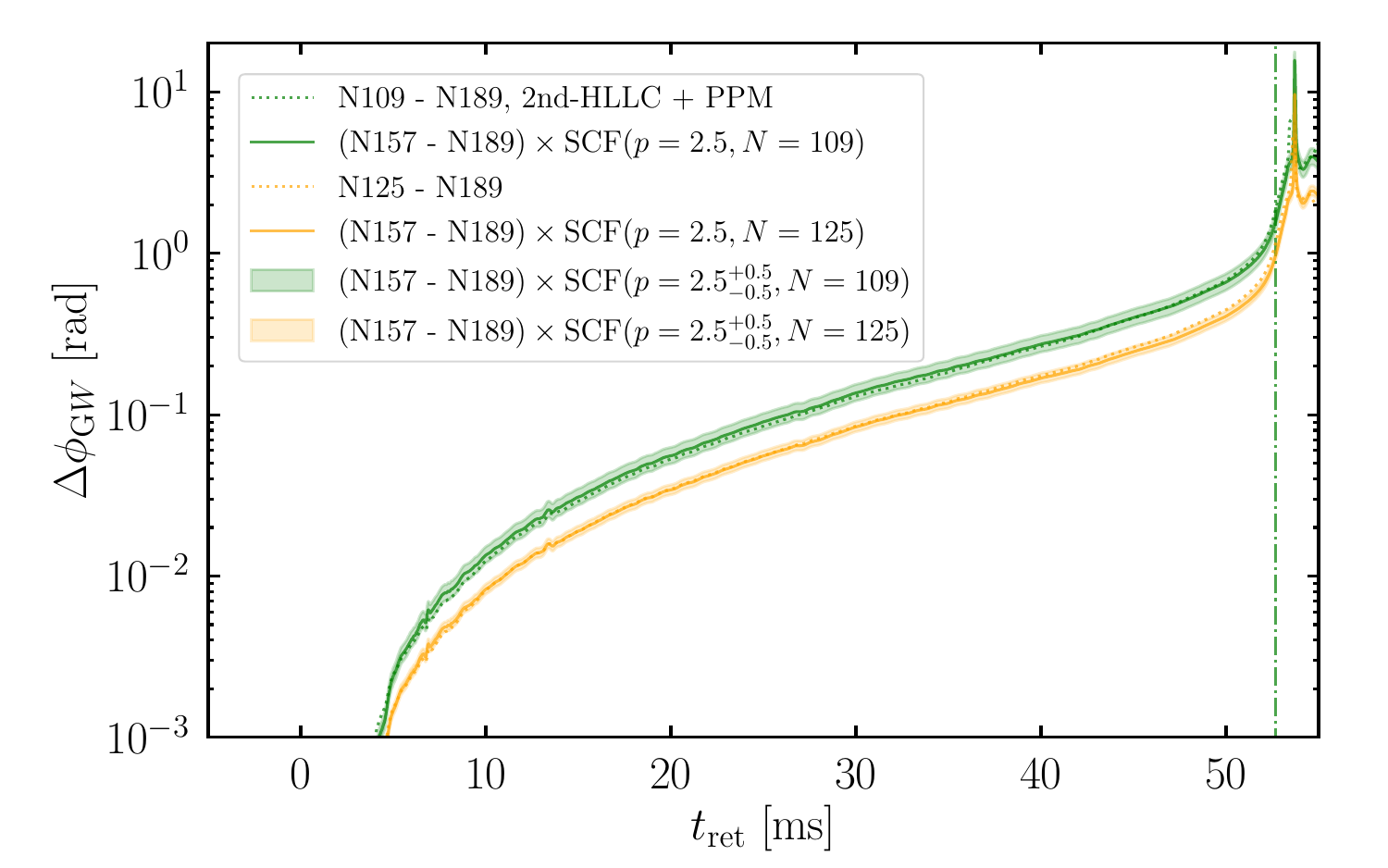}
    \caption{The gravitational phase error as a function of the reterded time for the second-order HLLC. The green- and orange-dashed curves show the gravitational wave phase error between $N=109$ and $189$, and between $N=125$ and $189$, respectively. The green- and orange-solid curves denote the phase error between $N=157$ and $189$ multiplied by the self-convergence factor assuming $p=2.5$, respectively. The color-shaded regions assume the convergence order with $2.5 \pm 0.5$.
    }
    \label{fig:sfcf}
\end{figure*}

Following Ref.~\cite{Dietrich:2019kaq}, we conduct a self-convergence test by calculating the self-convergence factor:
\begin{align}
{\rm SCF}(p,N) = \frac{\left(\Delta x_N/\Delta x_{189}\right)^p-1}{\left(\Delta x_{158}/\Delta x_{189}\right)^p-1},
\end{align}
where $N=109$ or $125$. $p$ denotes the convergence order. Figure~\ref{fig:sfcf} plots the phase error multiplied by the self-convergence factor with $p=2.5$ (the solid curves) estimated by the "human eye" on top of the phase error between $N=109$ and $189$ (the green-dotted curve) and between $N=125$ and $189$ (the orange-dotted curve) derived by the second-order HLLC. Strictly speaking, the two curves, i.e., the original phase error and error multiplied by the self-convergence factor, do not {\it perfectly} overlap during the entire inspiral phase, which is also the case in the literature~\cite{Dietrich:2019kaq,Bernuzzi:2015rla}. The color-shaded regions denote the phase error multiplied by the self-convergence factor, with the convergence order varying by $\pm 0.5$ from $2.5$, which also looks consistent with the original phase error. It demonstrates that the "human eye's inspection" of the convergence order in the self-convergence study is not stringent in the sense that it could overestimate/underestimate the convergence order. 

\bibliography{reference}

\end{document}